\documentclass[preprint,aps]{revtex4}
\usepackage{amssymb,amsmath}
\usepackage{graphics}
\usepackage{dcolumn}
\usepackage{bm}

\begin{document}

\title{Polyakov loops, Gross-Witten like point and Hagedorn states}
\author{
I. Zakout and C. Greiner}
\affiliation{
Institut f\"ur Theoretische Physik, J. W. Goethe-Universit\"at, \\
D-60438 Frankfurt am Main, Germany
}


\date{\today} 

\begin{abstract}
The phase transition for a finite volume system that incorporates 
the Polyakov loops and maintains the colorless state 
is explored using the Polyakov-loop extended 
Nambu-Jona-Lasinio (PNJL) model. 
The order parameter for Polyakov loops 
is demonstrated to signal the appearance 
of a transition for $SU(3)_{c}$ 
analogous to Gross-Witten (GW-) phase transition 
instead of the deconfinement phase transition 
to quark-gluon plasma. 
The asymptotic restoration of Polyakov loops 
is conjectured to be a threshold production 
for meta-stable Hagedorn (or semi-QGP) states
and this does not imply a direct deconfinement 
phase transition.
In this context, the GW-like point is the point 
where the colorless states switches 
from the low-lying hadronic states 
to the meta-stable high-lying Hagedorn states.
The chiral phase transition takes place within 
an extended GW-like point depending on the fireball's size.
The deconfinement phase transition is determined 
by Hagedorn's temperature above GW-like temperature.
\end{abstract}
\maketitle

\section{Introduction}
Recently, Fukushima ~\cite{Fukushima:2003fw} has extended 
the Nambu-Jona-Lasinio (NJL) model to include
Polyakov loops, namely, $\Phi$ and $\overline{\Phi}$
and the $\sigma$-chiral field.  
Fukushima's approach is known 
as Polyakov extended Nambu-Jona-Lasinio (PNJL) model.
The Polyakov loops are related to the imposition of the Gauss' law
where the trivial vacuum is a minimum of the free energy
and the formation of stable
colorless QG-droplet~\cite{Gocksch:1993iy}.
The PNJL model has been widely adopted to study 
the phase transition diagram. 
Furthermore, it has been extended to investigate 
the phase transition diagram with 
various phenomenological effective Polyakov
and gluon potentials as well as various extensions 
to include other NJL's fields such as 
the isospin scalar and vector fields 
as well as the color superconductivity
~\cite{Fukushima:2003fw,Fukushima:2008wg,
Ratti:2005jh,Roessner:2006xn,Sasaki:2010jz,Sakai:2008py,
Schaefer:2009ui,Schaefer:2007pw,Ciminale:2007sr,
Abuki:2008nm,Abuki:2008ht,Abuki:2008iv}.
The hybrid description of PNJL with two and three flavors 
has been studied 
extensively with various modifications 
of the effective Polyakov and gluon potential 
and the results have been compared  with lattice QCD 
data~\cite{Ratti:2005jh}.
The Polyakov potential as a function of Polyakov loops 
(or equivalently the VanderMonde potential as a function 
of fundamental eigenvalues of the Gauss-law)
is originated from the invariance Haar measure 
of $SU(N_{c})$ in order to project the colorless state 
of the quark and gluon fireball.
The comparison with lattice calculations 
~\cite{Ratti:2005jh,Schaefer:2009ui,Sasaki:2010jz,Fukushima:2008wg} 
hints that the effective Polyakov potential could 
by modified by temperature. 
Furthermore, it has been suggested that modifying
the invariance Haar measure's exponent somehow mutates
the Hagedorn's internal structure ~\cite{Zakout:2010ep}. 
Furthermore, the bag's volume fluctuation can modify
the effective Polyakov potential.
Nevertheless, the modification of the effective 
Polyakov potential in the medium and its impacts in the phase transition
diagram will be considered in a future work. 
The internal structure of the quark-gluon (QG)
has been suggested to be crucial 
to the tri-critical point 
(see for instance Ref.~\cite{Andronic:2009gj}
and reference therein). 
This has a significant impact in the recent research to explore 
the width of phase transition and intermediate processes 
such as Hagedorn states, quarkyonic matter 
and semi-classical QGP phases 
and color-flavor superconductor matter.
There exist various reviews discussing 
the QG-blob's internal color structure.
Brezin, Itzykson, Parisi and Zuber studied 
the planar approximation to field theory through the
limit of a large internal symmetry group ~\cite{Brezin:1977sv}. 
This procedure is known as the matrix saddle point method.
Gross and Witten ~\cite{Gross:1980he} using the matrix method 
have discovered a possible transition 
from a specific phase with strong coupling to another phase 
with weak coupling in the large $N_{c}$ limit 
(i.e. $N_{c}\,\rightarrow\,\infty$ but a finite $g^2\,N_{c}$)
of Wilson lattice gauge theory. 
For technical reasons, the spectral density method 
which has been developed by Brezin {\it et. al.}
~\cite{Brezin:1977sv,Gross:1980he} 
depends basically on 
the large $N_{c}$ limit and it is not permissible 
for technical reasons to extend the same analyses 
using the spectral density for finite number of colors. 
The GW-like point sticks in one's mind 
for $N_{c}\,\rightarrow\,\infty$ and remains obsolete 
for $N_{c}=3$ (i.e. the QCD). 
The GW-like phase transition for finite $N_{c}$
is not expected to have the same characteristic 
behavior to that one in the limit 
$N_{c}\,\rightarrow\,\infty$.
In order to search for a mechanism analogous to GW-transition
in QCD,  
it is important 
to extend the analysis using 
the (non-Gaussian-) stationary points method in the strong coupling limit 
in the context of the Polyakov loop parameterization as done 
by in Ref~\cite{Fukushima:2003fw} and the references therein 
on one hand and the (Gaussian-) saddle points approximation
in the weak coupling limit
as done by Elze, Greiner and Rafelski and others
~\cite{Elze:1986db,Elze:1986gz,Elze:1983du,Elze:1984un,
Elze:1985wv,Spieles:1997ab,
Zakout:2006zj,Zakout:2007nb,Gorenstein:1983a,Auberson:1986a,Mehta1967}
on the other hand.
The interpolation between the asymptotic 
non-Gaussian stationary points approximation's solution for the strong coupling limit 
and the asymptotic Gaussian saddle points approximation's solution 
for the weak coupling limit is not fully understood in QCD 
and the corresponding mechanism is analogous to GW-transition.
Furthermore, GW-like transition sounds to take place over 
the interpolation range between two asymptotic solutions 
(i.e. over an extended interval)
rather than a single deflection point.
Furthermore, Elze, Greiner and Rafelski have pointed out 
that the non-perturbative effect of the colorless state
leads to a gradual freezing of internal degrees of 
freedom ~\cite{Elze:1983du,Elze:1984un}.
This mechanism could explain
the emergence of QG liquid droplet(s) or equivalent forms such 
as Hagedorn states, quarkyonic droplets etc.
It should be stressed that GW-like transition 
is not a confinement/deconfinement phase transition, 
but instead is the production threshold of 
(meta-) Hagedorn states in hadronic matter. 
The Hagedorn states emerge as gas of bags. 
Therefore, there is a possibility for a new form 
of matter that can be formed in a narrow range above GW-like point
and below Hagedorn's temperature.
This form of matter emerges as a gas/liquid of bags 
and these bags expand and grow up gradually. 
When Hagedorn's temperature is reached, the system undergoes
a deconfinement phase transition to QGP.
 
The outline of the present paper is as follows: 
In Sec. II, we review Polyakov loops without chiral field 
and demonstrate the interpolation between the low-lying and high-lying 
energy solutions 
and a possible transition that is analogous to GW-transition.
In Sec. III, the treatment is extended to include 
the $\sigma$-chiral field 
in the context of PNJL model and demonstrate the emergence 
of an extended GW-like point. 
The connection between GW-like point and production of Hagedorn states 
is discussed in Sect. IV.
Finally, the conclusion is presented in Sec. V. 

\section{A simple canonical ensemble with Polyakov loops\label{sectionII}}

The grand potential for the quark and the anti-quark is given by
\begin{eqnarray}
\frac{
\Omega_{q\overline{q}}\left(\beta,V;\theta_{1},\theta_{2}\right)
}{V}
&=&
-\frac{1}{V\beta}\,\log_{e} 
Z_{q\overline{q}}\left(\beta,V;\theta_{1},\theta_{2}\right),
\nonumber\\
&=&-(2J+1) \sum^{N_f}_{q}\int \frac{d^{3} \vec{p}}{(2\pi)^{3}}
\sum^{N_{c}}_{i}\left[
\epsilon_{q}\left(\vec{p}\right)
+\frac{1}{\beta} \log_{e}
\left(1+e^{-\beta\left[\epsilon_{q}\left(\vec{p}\right)
-\mu_{q}-i\frac{\theta_{i}}{\beta}\right]}\right)
\right.
\nonumber\\
&~& ~~~ ~~~ ~~~ ~~~~
\left.
+\frac{1}{\beta} \log_{e}
\left(1+e^{-\beta\left[\epsilon_{q}\left(\vec{p}\right)
+\mu_{q}+i\frac{\theta_{i}}{\beta}\right]}\right)
\right],
\label{quark-antiquark-ens1}
\end{eqnarray}
where $\epsilon_{q}\left(\vec{p}\right)=\sqrt{\vec{p}^{2}+m^{2}_{q}}$, $(2J+1)=2$ 
is the spin degeneracy, $V$ is the quark and gluon blob's volume and
$\mu_{q}$ is the flavor chemical potential 
while $\theta_{i}$ are the imaginary color chemical potentials
or fundamental gauge fields of the $SU(N_{c})$ group's  fundamental representation
(i.e. Gauss-law's eigenvalues on the thermal excitations).
When no chiral fields are involved in the calculation, 
$m_{q}$ is reduced to the current mass 
(for only the sake of simplicity, 
it can be assumed massless for light flavors).
The first term in the square bracket that appears 
on the right hand side of Eq.(\ref{quark-antiquark-ens1})
is temperature independent.  It diverges at zero temperature and  
is a non-re-normalizable term. 
It can be regulated in the standard way by introducing 
UV-cutoff for the momentum integration.
In the standard $\sigma$-model, 
that term is trivially dropped as far it can be absorbed by the nonlinear 
$\sigma$-potential
but the this is not the case in Nambu-Jona-Lasinio model (NJL) 
where the first term is regularized and retained in the calculation.
After a simple algebraic manipulation, Eq.(\ref{quark-antiquark-ens1})
becomes
\begin{eqnarray}
\frac{\Omega_{q\overline{q}}
\left(\beta,V;\Phi,\overline{\Phi}\right)}{V}
&=&
\frac{
\Omega_{q\overline{q}}
\left(\beta,V;\theta_{1},\theta_{2}\right)}{V},
\nonumber\\
&=&
-2 N_{c}\sum^{N_f}_{q} \int^{\Lambda}_{0} \frac{d|\vec{p}| |\vec{p}|^{2}}{2{\pi}^{2}}
\epsilon_{q}\left(\vec{p}\right)
\nonumber\\
&~&
-\frac{2}{\beta} \sum^{N_f}_{q}\int \frac{d^{3} \vec{p}}{(2\pi)^{3}}
\nonumber\\
&~&  
\times
\left(
\log_{e}
\left[1+
3\left( \Phi+\overline{\Phi} e^{-\beta\left[\epsilon_{q}\left(\vec{p}\right)-\mu_{q}\right]}
\right)  e^{-\beta\left[\epsilon_{q}\left(\vec{p}\right)-\mu_{q}\right]}
+  e^{-3\beta\left[\epsilon_{q}\left(\vec{p}\right)-\mu_{q}\right]}
\right]
\right.
\nonumber\\
&~& ~
\left.
+
\log_{e}
\left[1+
3\left( \overline{\Phi}+\Phi e^{-\beta\left[\epsilon_{q}\left(\vec{p}\right)+\mu_{q}\right]}
\right)  e^{-\beta\left[\epsilon_{q}\left(\vec{p}\right)+ \mu_{q}\right]}
+  e^{-3\beta\left[\epsilon_{q}\left(\vec{p}\right)+\mu_{q}\right]}
\right]
\right),
\label{quark-antiquark-ens2a}
\end{eqnarray}
where $\Lambda$ is UV-cutoff that regularizing 
the divergent term over the momentum integration.
The UV-cutoff for momentum integration is taken 
$\Lambda=631.5$ MeV in the present calculations.
The Polyakov-loop triality parameters 
$\Phi$ and $\overline{\Phi}$ 
are defined, respectively, as follows
\begin{eqnarray}
\Phi&=&\frac{1}{N_{c}}\left[e^{i\theta_{1}}+e^{i\theta_{2}}+e^{i\theta_{3}}\right],
\nonumber\\
\overline{\Phi}&=&\frac{1}{N_{c}}\left[e^{-i\theta_{1}}+e^{-i\theta_{2}}+e^{-i\theta_{3}}\right],
\label{Polyakov-loops-1}
\end{eqnarray}
where $\theta_{3}=-\theta_{1}-\theta_{2}$ for $SU(3)_{c}$ 
and the fundamental gauge fields $\theta_{i},i=1,2,3$ 
are subjected to the periodicity condition over the interval 
$-\pi\le \theta_{i}\le \pi$.
Eq.(\ref{quark-antiquark-ens2a}) can be written as follows
\begin{eqnarray}
\frac{\Omega_{q\overline{q}}\left(\beta,V;\Phi,\overline{\Phi}\right)}
{V}
&=&
-2 N_{c}\sum^{N_f}_{q} \int^{\Lambda}_{0} \frac{d|\vec{p}| |\vec{p}|^{2}}{2{\pi}^{2}}
\epsilon_{q}\left(\vec{p}\right)
\nonumber\\
&~&
- 6 \sum^{N_f}_{q}\int \frac{d |\vec{p}|}{(2\pi^{2})} \frac{|\vec{p}|^{4}}
{3 \epsilon_{q}\left(\vec{p}\right)}
\nonumber\\
&~& ~~~ ~~~ 
\times
\left(
\frac{
\left( \Phi+2\overline{\Phi} e^{-\beta\left[\epsilon_{q}\left(\vec{p}\right)-\mu_{q}\right]}
\right)  e^{-\beta\left[\epsilon_{q}\left(\vec{p}\right)-\mu_{q}\right]}
+  e^{-3\beta\left[\epsilon_{q}\left(\vec{p}\right)-\mu_{q}\right]}
}{
1+
3\left( \Phi+\overline{\Phi} e^{-\beta\left[\epsilon_{q}\left(\vec{p}\right)-\mu_{q}\right]}
\right)  e^{-\beta\left[\epsilon_{q}\left(\vec{p}\right)-\mu_{q}\right]}
+  e^{-3\beta\left[\epsilon_{q}\left(\vec{p}\right)-\mu_{q}\right]}
}
\right.
\nonumber\\
&~& ~~~ ~~~ ~~~ 
+
\left.
\frac{
\left( \overline{\Phi}+2\Phi e^{-\beta\left[\epsilon_{q}\left(\vec{p}\right)+\mu_{q}\right]}
\right)  e^{-\beta\left[\epsilon_{q}\left(\vec{p}\right)+\mu_{q}\right]}
+  e^{-3\beta\left[\epsilon_{q}\left(\vec{p}\right)+\mu_{q}\right]}
}{
1+
3\left( \overline{\Phi} + \Phi e^{-\beta\left[\epsilon_{q}\left(\vec{p}\right)+\mu_{q}\right]}
\right)  e^{-\beta\left[\epsilon_{q}\left(\vec{p}\right)+\mu_{q}\right]}
+  e^{-3\beta\left[\epsilon_{q}\left(\vec{p}\right)+\mu_{q}\right]}
}
\right).
\label{quark-antiquark-ens2}
\end{eqnarray}
In the case of massless flavors and $\mu_{q}=0$ 
and in the terms of fundamental gauge fields, 
Eq.(\ref{quark-antiquark-ens2}) reads 
\begin{eqnarray}
-\frac{\Omega_{q\overline{q}}\left(\beta,V;\Phi,\overline{\Phi}\right)}{V}
&=&
-\frac{\Omega_{q\overline{q}}\left( \beta,V;\Phi(\theta_{1},\theta_{2}),
\overline{\Phi}(\theta_{1},\theta_{2}) \right)}{V},
\nonumber\\
&=& -\frac{\Omega_{q\overline{q}}\left(\beta,V;\theta_{1},\theta_{2})\right)}{V},
\nonumber\\
&=&
\frac{\Lambda^{4}}{4\pi^{2}}\, N_{f} N_{c}
+ \frac{7 \pi^{2}}{180 \beta^{4}} \,N_{f} N_{c}\,
- \frac{1}{6 \beta^{4}}\,N_{f}\,\sum^{N_{c}}_{i=1}
\theta_{i}^{2}\left(1-\frac{\theta_{i}^{2}}{2\pi^{2}}\right).
\label{quark-antiquark-eigen-ens2}
\end{eqnarray}
The canonical ensemble for the quark and anti-quark becomes
\begin{eqnarray}
Z_{q\overline{q}}\left(\beta,V;\Phi,\overline{\Phi}\right)&=&
\exp\left[ -\beta\, \Re e\,
\Omega_{q\overline{q}}\left(\beta,V;\Phi,\overline{\Phi}\right) 
\right],
\end{eqnarray}
where 
$\Omega_{q\overline{q}}\left(\beta,V;\Phi,\overline{\Phi}\right)$
is given by Eq.(\ref{quark-antiquark-ens2}).
Fortunately, the $q\overline{q}$ grand potential becomes 
a real one when $\mu_{q}=0$. 
The partition function for the gluons can be calculated in a similar manner. 
The phenomenological gluon potential parameterized in the terms of 
Polyakov loops has been adopted recently 
in the literature ~\cite{Fukushima:2003fw}.
The general choice is given by
\begin{eqnarray}
\frac{1}{V}\Omega_{g}\left(\beta,V;\Phi,\overline{\Phi}\right)&=&
-\frac{1}{V}
\frac{1}{\beta} \log_{e} Z_{g}\left(\beta,V;\Phi,\overline{\Phi}\right),
\nonumber\\
&=&
- 2\frac{1}{\beta^{4}}\left( \frac{a(T)}{4} \right) \overline{\Phi} \Phi,
\label{phen-gluon-partition1f}
\end{eqnarray}
where Polyakov term $\overline{\Phi} \Phi$
can be written in the terms of fundamental gauge fields as follows
\begin{eqnarray}
\overline{\Phi} \Phi= \frac{1}{N_{c}^{2}}
\sum^{N_{c}}_{i,j=1} \cos\left(\theta_{i}-\theta_{j}\right),
\end{eqnarray}
and 
\begin{eqnarray}
a(T)=a_{0}+a_{1} \left(\frac{T_{0}}{T}\right) + a_{2} \left(\frac{T_{0}}{T}\right)^{2},
\end{eqnarray}
where $a_{0}$, $a_{1}$ and $a_{2}$ are phenomenological parameters. 
The phenomenological gluon partition function is usually adopted 
in the term of Polyakov loop approach in order to replace the standard 
gluon partition function that is given by
\begin{eqnarray}
Z_{g}\left(\beta,V;\theta_{1},\theta_{2}\right)&=&\exp\left[
-2 V \int \frac{d^{3} \vec{p}}{(2\pi)^{3}}
\sum^{N_{c}^{2}-1}_{a=1} 
\log_{e}\left(
1-e^{ -\left(\beta\, \epsilon_{g}(\vec{p})-i\phi^{a}\right) } 
\right)
\right],
\nonumber\\
&=&\exp\left[
-2 V \int \frac{d^{3} \vec{p}}{(2\pi)^{3}}
\sum^{N_{c}}_{i} \sum^{N_{c}}_{j}
\log_{e}\left(
1-e^{ -\left(\beta\, \epsilon_{g}(\vec{p})-i(\theta_{i}-\theta_{j})\right) }
\right) 
\right],
\label{gluon-partition-1}
\end{eqnarray}
where $\epsilon_{g}(\vec{p})=|\vec{p}|$. 
Eq.(\ref{gluon-partition-1}) is evaluated explicitly as follows
\begin{eqnarray}
\log_{e} Z_{g}\left(\beta,V;\theta_{1},\theta_{2}\right)&=&
\frac{2V}{\beta^{3}}\left[
\left(N^{2}_{c}-1\right) \frac{\pi^{2}}{90}
-\frac{1}{6}\sum^{N_{c}}_{i<j} \left(\theta_{i}-\theta_{j}\right)^{2}
\left(
1-\frac{\left|\theta_{i}-\theta_{j}\right|}{2\pi}
\right)^{2}
\right].
\label{gluon-partition-explicit-1}
\end{eqnarray}
In the standard treatment the gluons are treated 
as the adjoint interaction particles of the $SU(N_{c})$ 
symmetry group.
It should be noted here that 
the $SU(N_{c})$'s adjoint eigenvalues (i.e. adjoint gauge fields), 
namely, $\phi^{a}$ 
are calculated from the nested commutation relations 
for the fundamental eigenvalues, namely, $\theta_{i}$ of the Lie algebra.
The adjoint eigenvalues are related to fundamental eigenvalues by the relation 
$\phi^{a}\equiv \left(\theta_{i}-\theta_{j}\right)$. 
This relation diagonalizes the adjoint representation 
and subsequently commutes with the Hamiltonian.
In order to understand the origin of the gluon's phenomenological potential,
Eq.(\ref{gluon-partition-1}) can be approximated and simplified in order to be evaluated 
using Polyakov loop variables in the following systematic way
\begin{eqnarray}
Z_{g}\left(\beta,V;\Phi,\overline{\Phi}\right)
&\approx&\exp\left[
2 V \beta \sum^{N_{c}^{2}-1}_{a=1}
\int \frac{d |\vec{p}| }{2\pi^{2}} \frac{|\vec{p}|^{3}}{3}
\frac{1}{ e^{i\phi^{a}} e^{\beta |\vec{p}|}-1 }
\right],
\nonumber\\
&=&\exp\left[
-2 V \beta        
\int \frac{d |\vec{p}| }{2\pi^{2}} \frac{|\vec{p}|^{3}}{3}
\frac{\sum^{8}_{n=1} n\,C_{n}\,e^{-n \beta |\vec{p}|}}
{1+\sum^{8}_{n=1} C_{n}\,e^{-n \beta |\vec{p}|} }
\right],   
\label{gluon-partition-approx-2-prior}
\end{eqnarray}
where the factor $2$ that appears on the right hand side
comes from the spin degeneracy.
The coefficients $C_{n}$ are functions of Polyakov loops
\footnote{This point has been brought to the attention of the authors by C. Sasaki}.
When the Polyakov loops vanish ($\Phi,\overline{\Phi}\rightarrow 0$), 
the gluon grand potential is reduced to 
$\lim_{\Phi,\overline{\Phi}\rightarrow 0}
\frac{1}{V} \Omega_{g}
\left(\beta,V;\Phi,\overline{\Phi}\right) = -\frac{1}{\beta^{4}}
\frac{1}{N_{c}^{2}} 
\frac{\left(N_{c}^{2}-1\right) \pi^{2}}{45}$,
while in the case of Polyakov loop restoration 
($\Phi,\overline{\Phi}\rightarrow 1$), it is reduced to
$\lim_{\Phi,\overline{\Phi}\rightarrow 1}
\frac{1}{V} \Omega_{g}
\left(\beta,V;\Phi,\overline{\Phi}\right) = -\frac{1}{\beta^{4}}
\frac{\left(N_{c}^{2}-1\right) \pi^{2}}{45}$.
This implies that $\Omega_{g}$ is reduced by factor $1/N_{c}^{2}$ 
when $\Phi,\overline{\Phi}$ are changed from 1 to 0. 
It can be parameterized to 
$\frac{1}{V} \Omega_{g}
\left(\beta,V;\Phi,\overline{\Phi}\right) \approx -\frac{1}{\beta^{4}}
\frac{\left[\left(N_{c}^{2}-1\right)\Phi\overline{\Phi}+1\right]}
{N_{c}^{2}} 
\frac{\left(N_{c}^{2}-1\right)\pi^{2}}{45}$.
%
%
%
%
%
%
%
%
%
%
%
%
%
%
%
%
%
In order to simplify the calculation drastically, 
the gluon grand potential
is simplified to a phenomenological potential such as 
that one given in Eq.(\ref{phen-gluon-partition1f}) as follows
\begin{eqnarray}
\frac{1}{V} \Omega_{g}\left(\beta,V;\Phi,\overline{\Phi}\right)
& \equiv & - \frac{1}{\beta^{4}}\, {\omega}_{g}\,\overline{\Phi} \Phi,
\nonumber\\
&=&
- \frac{1}{\beta^{4}}\, {\omega}_{g}\,
\frac{1}{ N^{2}_{c} }
\,\left[
3+2\sum^{N_c}_{i<j}\cos\left(\theta_{i}-\theta_{j}\right)
\right],
\label{phen-gluon-partition1}
\end{eqnarray}
where 
\begin{eqnarray}
\omega_{g}=\left(N_{c}^{2}-1\right) \frac{\pi^{2}}{45}.
\end{eqnarray}
In the calculation of the phase transition from the low-lying energy 
excitations to the high-lying ones 
but below the deconfinement phase transition, 
it is adequate to use the potential that is given 
by Eq.(\ref{phen-gluon-partition1}).

The canonical ensemble for a finite volume quark and gluon blob 
in the Hilbert space is given by the Fock product of quark and antiquark 
partition function and the gluon partition function as follows
\begin{eqnarray}
Z_{q\overline{q} g}\left(\beta,V;\theta_{1},\theta_{2}\right)&=&
Z_{q\overline{q} g}
\left(\beta,V;\theta_{1},\theta_{2},\theta_{3}=-\theta_{1}-\theta_{2}\right),
\nonumber\\
&=&
Z_{q\overline{q} g}\left(\beta,V;\Phi(\theta_{1},\theta_{1}),
\overline{\Phi}(\theta_{1},\theta_{1})\right),
\nonumber\\
&=&
Z_{q\overline{q}}\left(\beta,V;\Phi,\overline{\Phi}\right)
\times Z_{g}\left(\beta,V;\Phi,\overline{\Phi}\right).
\end{eqnarray}
This implies that the grand canonical ensemble is reduced to 
\begin{eqnarray}
Z_{q\overline{q} g}(\beta,V;\Phi,\overline{\Phi})&=& \exp\left(
-\beta 
\Omega_{q\overline{q} g}\left(\beta,V;\Phi,\overline{\Phi}\right)
\right),
\nonumber\\
&=& \exp\left(
-\beta 
\Omega_{q\overline{q} g}\left(\beta,V;\theta_{1},\theta_{2}\right)
\right),
\end{eqnarray}
where 
\begin{eqnarray}
\Omega_{q\overline{q} g}\left(\beta,V;\Phi,\overline{\Phi}\right)
= 
\Re e \Omega_{q\overline{q}}\left(\beta,V;\Phi,\overline{\Phi}\right)
+\Omega_{g}\left(\beta,V;\Phi,\overline{\Phi}\right).
\end{eqnarray}
The colorless state for the quark and gluon blob 
is ensured by projecting 
the color singlet state in the following way
\begin{eqnarray}
Z_{colorless}\left(\beta,V\right)&=&
\int d\mu\left({\bf g}\right)\,
e^{\beta\,\Omega_{q\overline{q} g}\left(\beta,V;\Phi,\overline{\Phi}\right)},
\nonumber\\
&=&
\frac{1}{N!}
\prod^{N_{c}-1}_{k=1} 
\left( \int^{\pi}_{-\pi}\frac{d \theta_{k}}{2\pi} \right)
\,
e^{-\beta {\cal V}_{VdM}\left({\bf g}\right) }
\,
Z_{q\overline{q}g}
\left(\beta,V;\Phi,\overline{\Phi}\right),
\nonumber\\
&=&
\frac{1}{N!}
\prod^{N_{c}-1}_{k=1}
\left( \int^{\pi}_{-\pi}\frac{d \theta_{k}}{2\pi} \right)
\,
e^{-\beta\left[ 
{\cal V}_{VdM}\left({\bf g}\right) 
+
\Omega_{q\overline{q} g}\left(\beta,V;\Phi,\overline{\Phi}\right)
\right]
},
\label{ex-colorless-projection}
\end{eqnarray}
where
\begin{eqnarray}
\int d \mu({\bf g})&=&
\frac{1}{N_{c}!}\frac{1}{(2\pi)^{2}}
\int^{\pi}_{-\pi} d\theta_{1}
\int^{\pi}_{-\pi} d\theta_{2}
\int^{\pi}_{-\pi} d\theta_{3}
\delta\left(\sum^{3_{c}}_{i=1}\theta_{i}\right)
\,
\prod_{i<j} \left|2 \sin\left(\frac{\theta_{i}-\theta_{j}}{2}\right)\right|^{2},
\nonumber\\
&=&
\frac{1}{N_{c}!}\frac{1}{(2\pi)^{2}}
\int^{\pi}_{-\pi} d\theta_{1}
\int^{\pi}_{-\pi} d\theta_{2}
\,
\prod_{i<j} \left|2 \sin\left(\frac{\theta_{i}-\theta_{j}}{2}\right)\right|^{2}.
\label{invriance-Haar-measure-eigen1}
\end{eqnarray}
The VanderMonde potential is stemmed 
from the invariance Haar measure of the group integration
and is defined by
\begin{eqnarray}
{\cal V}_{VdM}\left({\bf g}\right)&=&
-\frac{1}{\beta}\,{G}_{sym}\sum^{N_{c}}_{i<j}\log_{e}
\left[2\sin\left(\frac{\theta_{i}-\theta_{j}}{2}\right)\right],
\end{eqnarray}
where the parameter ${G}_{sym}$ 
depends basically on the group's symmetry.
It is reduced to ${G}_{sym}=$ 2 for $SU(N_{c})$.
In the lattice modeling, the number of states
for the VanderMonde potential~\cite{Gocksch:1993iy}
~\footnote{This point has been brought to
the attention of the authors by R. Pisarski.} 
is introduced by 
$\int d^{d} x \delta^{d}\left(0\right)$
as follows
\begin{eqnarray}
\sum\left(\mbox{states}\right)\,&\rightarrow&\, 
\frac{1}{a^{3}}\int dV,
\nonumber\\
&\rightarrow&\,
\frac{V}{a^{3}},
\end{eqnarray}
where $a^{3}$ is the lattice size. 
Hence, the VanderMonde potential is regulated~\cite{Gocksch:1993iy} 
as follows
\begin{eqnarray}
{\cal V}_{VdM}\left({\bf g}\right)
&=&
-\frac{1}{\beta}\,{G}_{sym}
\,\left(\frac{V}{ a^{3} }\right)\,
\sum^{N_{c}}_{i<j}\log_{e}
\left[2\sin\left(\frac{\theta_{i}-\theta_{j}}{2}\right)\right].
\end{eqnarray}
A finite bag with volume 
at the same size order of the lattice
$V\sim a^{3} \sim \mbox{fm}^{3}$ 
and 
$\frac{1}{a^{3}} V\sim 1$
is considered in the present work.
The regulation $\gamma_{reg}=\frac{V}{a^{3}}$ 
will be considered elsewhere.
However, the term  $\frac{1}{a^{3}} V$  
in MIT bag model is related 
to the volume fluctuation for a bag with an extended surface.
Nonetheless, the VanderMonde potential regulation is essential 
for a system with infinite volume~\cite{Gocksch:1993iy}. 
Hereinafter, the number of states is considered
$\frac{1}{a^{3}} V\equiv 1$ for VanderMonde potential
in colorless quark and gluon bag.

In order to consider Polyakov loops parameterization, 
it is useful to perform the variable transformation from 
fundamental gauge fields, 
namely, $(\theta_{1},\theta_{2},\theta_{3})$ 
with $\theta_{3}=-\theta_{1}-\theta_{2}$
to Polyakov loop variables, namely, 
$\left(\Phi,\overline{\Phi}\right)$.
In the context of $SU(3)_{c}$, 
the invariance Haar measure is furnished by
\begin{eqnarray}
\int d \mu({\bf g})&=&
\frac{1}{N!}\frac{1}{(2\pi)^{2}} 
\int^{\pi}_{-\pi} d\theta_{1}\, \int^{\pi}_{-\pi} d\theta_{2}\,
\prod_{i<j} \left|2 \sin\left(\frac{\theta_{i}-\theta_{j}}{2}\right)\right|^{2},
\nonumber\\
&=&
\frac{1}{N!}\frac{1}{(2\pi)^{2}} \int^{\pi}_{-\pi} d\theta_{1}\,
\int^{\pi}_{-\pi} d\theta_{2}\,
\left(
27\left[
1-6\Phi \overline{\Phi}+4\left(\Phi^{3}+\overline{\Phi}^{3}\right)-
3\left(\Phi \overline{\Phi}\right)^{2}
\right]\right).
\end{eqnarray}
The invariance Haar measure can be transformed and written 
in the terms 
of Polyakov loop variables $\Phi$ and $\overline{\Phi}$. 
The transformation of the square root of the invariance Haar measure
from the variable set $\{\theta_{i}\}$ 
to 
$\Phi$ and $\overline{\Phi}$
leads to
\begin{eqnarray}
\prod_{i<j} \left|2 \sin\left(\frac{\theta_{i}-\theta_{j}}{2}\right)\right|&=&
\left(
27\left[
1-6\Phi \overline{\Phi}+4\left(\Phi^{3}+\overline{\Phi}^{3}\right)-
3\left(\Phi \overline{\Phi}\right)^{2}
\right]\right)^{\frac{1}{2}}.
\end{eqnarray}
The integration over $\theta_{1}$ and  $\theta_{2}$ 
is transformed to Polyakov loop variables 
$\Phi$ and $\overline{\Phi}$ as follows
\begin{eqnarray}
\int\,d\theta_{1}\,d\theta_{2}
&=& \int d\Phi\, d\overline{\Phi} 
\left|
\frac{\partial\left(\Phi,\overline{\Phi}\right)}{\partial\left(\theta_{1},\theta_{2}\right)}
\right|^{-1},
\nonumber\\
&=&
 \int  d\Phi\, d\overline{\Phi} 
\left(
27\left[
1-6\Phi \overline{\Phi}+4\left(\Phi^{3}+\overline{\Phi}^{3}\right)-
3\left(\Phi \overline{\Phi}\right)^{2}
\right]\right)^{-\frac{1}{2}}.
\end{eqnarray}
Hence, the invariance Haar measure that is given by 
Eq.(\ref{invriance-Haar-measure-eigen1}) becomes
\begin{eqnarray}
\int d \mu({\bf g})&=&
N_{\mbox{Haar}}\,
\int_{C} d\Phi\, \int_{C} d\overline{\Phi}
\,
\left[
1-6\Phi \overline{\Phi}+4\left(\Phi^{3}+\overline{\Phi}^{3}\right)-
3\left(\Phi \overline{\Phi}\right)^{2}
\right]^{\frac{1}{2}},
\label{invariance-measure-Polyakov1and2}
\end{eqnarray} 
where $N_{\mbox{Haar}} =
\frac{ \sqrt{27} }{ {N_{c}}! \,(2\pi)^{{N_{c}}-1} }$ 
for $SU(3)_{c}$. 
The subscript notation, namely, $C$ that appears under the integral 
indicates the integration is over a complex plane domain.
The complex domain for $\Phi$ and $\overline{\Phi}$
is the three pointed star with a radius 1. 
The complex domain for Polyakov loops complicates 
the situation when the non-Gaussian stationary point method fails 
and the Gaussian saddle point procedure turns to be essential.
The invariance Haar measure with Polyakov loops parameterization
can be represented as
an effective Polyakov VanderMonde (PVdM) potential.
The colorless canonical ensemble
with an effective PVdM potential in $SU(3)_{c}$ 
group representation reads
\begin{eqnarray}
Z_{colorless} \left(\beta,V\right)
&=&
N_{\mbox{Haar}}
\int_{C} d\Phi\, \int_{C} d\overline{\Phi}
\left[
1-6\Phi \overline{\Phi}+4\left(\Phi^{3}+\overline{\Phi}^{3}\right)
-3\left(\Phi \overline{\Phi}\right)^{2} \right]^{\frac{1}{2}}
\nonumber\\
&~&
~\times\,
Z_{q\overline{q} g}\left(\beta,V;\Phi,\overline{\Phi}\right),
\nonumber\\
&=&
N_{\mbox{Haar}}
\int_{C} d\Phi\, \int_{C} d\overline{\Phi}
e^{-\beta 
\left[
{\cal V}_{PVdM} \left(\beta;\Phi,\overline{\Phi}\right)
+
\Omega_{q\overline{q} g}\left(\beta,V;\Phi,\overline{\Phi}\right)
\right]
},
\label{ensemble-Polyakov-Haar1}
\end{eqnarray}
where PVdM potential in $SU(3)_{c}$ representation
is given by
\begin{eqnarray}
{\cal V}_{PVdM}\left(\beta; \Phi, \overline{\Phi} \right)&=&
-\frac{1}{2} \frac{1}{\beta}
\log_{e}
\left(
1-6 \Phi \overline{\Phi} + 4 \left(\Phi^{3}+\overline{\Phi}^{3}\right)
- 3 \left(\Phi \overline{\Phi}\right)^{2}
\right),
\end{eqnarray}
and
\begin{eqnarray}
\Omega_{q\overline{q} g}\left(\beta,V;\Phi,\overline{\Phi}\right)&=&
-
\frac{1}{\beta}\,\log_{e}
\left[  
Z_{q\overline{q} g}\left(\beta,V;\Phi,\overline{\Phi}\right)
\right].
\label{qgp-colorless-ensemble-ex}
\end{eqnarray}
The phenomenological PVdM potential can be introduced by adding
a phenomenological pre-factor parameter, namely, $\alpha_{ph}$,
in front of the logarithm as follows
\begin{eqnarray}
{\cal V}_{PVdM}\left(\beta;\Phi ,\overline{\Phi} \right)
&=&
-\frac{1}{2} \frac{1}{\beta}
\,\alpha_{ph}\,
\log_{e}\left(
1-6 \Phi \overline{\Phi} + 4 \left(\Phi^{3}+\overline{\Phi}^{3}\right)
- 3 \left(\Phi \overline{\Phi}\right)^{2}
\right).
\end{eqnarray}
This phenomenological parameter, namely, $\alpha_{ph}$
modifies the underlying internal symmetry 
of Hagedorn states and in some scenarios 
this could break the internal symmetry
of the QG-bags but not the global symmetry
of the system~\cite{Zakout:2010ep}.
The phenomenological PVdM potential and the variation
of the phenomenological parameter, $\alpha_{ph}$,
will be considered in another work.
The integral that is given by Eq.(\ref{ensemble-Polyakov-Haar1}) 
is evaluated using the non-Gaussian stationary points method 
over the complex plane.
The Polyakov's stationary points, namely, $ \Phi=\Phi_{0}$ 
and $\overline{\Phi}=\overline{\Phi}_{0}$
are evaluated by extremizing 
the exponent term.
The stationary points $\Phi_{0}$ and $\overline{\Phi}_{0}$
are calculated as follows
\begin{eqnarray}
\begin{array}{c}
\frac{1}{V}
\left.\frac{\partial}{\partial \Phi}\left[
{\cal V}_{PVdM} \left(\beta;\Phi,\overline{\Phi}\right)
+
\Omega_{q\overline{q} g}\left(\beta,V;\Phi,\overline{\Phi}\right)
\right]
\right|_{\Phi=\Phi_{0},\overline{\Phi}=\overline{\Phi}_{0}} = 0,
\\
\frac{1}{V}
\left.
\frac{\partial}{ \partial \overline{\Phi} }\left[
{\cal V}_{PVdM} \left(\beta;\Phi,\overline{\Phi}\right)
+
\Omega_{q\overline{q} g}\left(\beta,V;\Phi,\overline{\Phi}\right)
\right]\right|_{\Phi=\Phi_{0},\overline{\Phi}=\overline{\Phi}_{0}} = 0.
\end{array}
\label{stationary-points-solution1}
\end{eqnarray}
The $\Phi$'s extremum is determined by the following constraint,
\begin{equation}
\begin{array}{l}
\frac{3}{V}
\frac{\left[\overline{\Phi} -2\Phi^{2}+\Phi \overline{\Phi}^{2}\right]}
{\left[
1-6\Phi\overline{\Phi}+4\left(\Phi^{3}+\overline{\Phi}^{3}\right)
-3\left(\Phi\overline{\Phi}\right)^{2}
\right]}
=
\\
~~~ ~~~ ~~~ ~~~ ~~~
2\sum^{N_{f}}_{q}\int \frac{p^{2}\,dp}{2\pi^{2}}
\frac{
3 e^{-\beta\left(\epsilon_{q}\left(\vec{p}\right)-\mu_{q}\right)}
}{
\left[
1+3\left(
\Phi+\overline{\Phi} e^{-\beta\left(\epsilon_{q}\left(\vec{p}\right)-\mu_{q}\right)}
\right) e^{-\beta\left(\epsilon_{q}\left(\vec{p}\right)-\mu_{q}\right)}
+
 e^{-3\beta\left(\epsilon_{q}\left(\vec{p}\right)-\mu_{q}\right)}
\right]}
\\
~~~ ~~~ ~~~ ~~~
+
2\sum^{N_f}_{q}\int \frac{p^{2}\,dp}{2\pi^{2}}
\frac{
3 e^{-2\beta\left(\epsilon_{q}\left(\vec{p}\right)+\mu_{q}\right)}
}{
\left[
1+3\left(
\overline{\Phi}+\Phi e^{-\beta\left(\epsilon_{q}\left(\vec{p}\right)+\mu_{q}\right)}
\right) e^{-\beta\left(\epsilon_{q}\left(\vec{p}\right)+\mu_{q}\right)}
+
 e^{-3\beta\left(\epsilon_{q}\left(\vec{p}\right)+\mu_{q}\right)}
\right]}
+ \omega_{g}\, T^{3}\, \overline{\Phi}.
\end{array}
\label{Polyakov-parameters-chiral-sol-1}
\end{equation}
The same thing can be done for $\overline{\Phi}$.
As far as the nuclear matter environment remains in
the circumstance that Polyakov's stationary points 
are located in the region 
$\Phi_{0}<1$ and $\overline{\Phi}_{0}<1$ 
(i.e. non-Gaussian stationary points) 
and below the threshold of GW-like phase transition, 
then the canonical ensemble which is given 
by Eq.(\ref{ensemble-Polyakov-Haar1}), 
is evaluated as follows
\begin{eqnarray}
Z^{(I)}_{colorless}\left(\beta,V\right)&=&
\exp\left(
-
\beta\, \left[
{\cal V}_{PVdM}\left(\beta;\Phi_{0},\overline{\Phi}_{0}\right)
+
\Omega_{q\overline{q} g}\left(\beta,V;\Phi_{0},\overline{\Phi}_{0}\right)
\right]
\right).
\label{ensemble-Polyakov-Haar-stationary1}
\end{eqnarray}
The pre-factor constant, namely $N_{\mbox{Haar}}$,
that appears in Eq.(\ref{ensemble-Polyakov-Haar1}) 
is dropped in order to normalize the partition function.  
The solution that is given by 
Eq.(\ref{ensemble-Polyakov-Haar-stationary1})
is assigned as the low-lying energy solution (I). 
This solution is the asymptotic solution below the threshold
of GW-like phase transition point. 
The validity of the low-lying energy solution (I)
is satisfied as far Polyakov's non-Gaussian 
stationary points remain in the energy domain 
$\Phi_{0}<<1$ and $\overline{\Phi}_{0}<<1$
(i.e. far away from Polyakov triality restoration point).
In the case that $\mu_{q}=0$, then the equations' set given 
by Eq.(\ref{Polyakov-parameters-chiral-sol-1}) 
becomes symmetry over Polyakov loop variables 
and this leads to equal and real stationary points
for $\Phi$ and $\overline{\Phi}$.  
However, whenever $|\Phi|_{0}\rightarrow 1^{-}$, 
then the effective PVdM potential, namely 
${\cal V}_{PVdM}\left(\beta;\Phi,\overline{\Phi}\right)$,
develops a virtual logarithmic divergence. 
Therefore, the logarithmic divergence of the effective PVdM 
potential spoils badly Polyakov's 
non-Gaussian stationary points procedure 
and leads to virtual singularity 
for the effective grand potential of the system.
Evidently, this virtual singularity deforms 
the low-lying energy solution
$Z^{(I)}_{colorless}\left(\beta,V\right)$. 
This kind of behavior indicates modification 
in the analytic behavior of the canonical ensemble 
and another analytical solution, 
namely the solution (II) emerges in the system. 
The change in the analytical solution is 
the beneath mechanism of GW-like phase transition
even for finite number of colors (i.e. $N_{c}=3$).
The GW-like point for $N_{c}=3$ may play 
a significant role in the deconfinement phase transition 
diagram in nuclear physics
as far it is not a confinement/deconfinement point.  
At the onset of GW-like phase transition,
the non-Gaussian stationary points turn 
to behave as Gaussian saddle points that
oscillate harmonically around the stationary points. 
This mechanism reflects the modification in the analytical 
behavior from the asymptotic solution (I) 
to the solution (II) when the temperature reaches GW-like point.  
Therefore, when GW-like threshold is reached, 
the canonical ensemble modifies its 
characteristic behavior from the low-lying 
energy solution
$Z^{(I)}_{colorless}\left(\beta,V\right)$ 
to the high-lying energy solution 
$Z^{(II)}_{colorless}\left(\beta,V\right)$.
The second solution implies a possible production of Hagedorn states.
When the asymptotic high-lying energy solution 
( i.e. solution (II) ) is reached, 
it becomes more suitable to write the invariance Haar measure 
in the terms of fundamental gauge fields 
(i.e. $\theta_{1}$, $\theta_{2}$, $\theta_{3}=-\theta_{1}-\theta_{2}$)
rather than Polyakov loop variables 
(i.e. $\Phi,\overline{\Phi}$).
Therefore, at the threshold of GW-like phase transition, 
the colorless canonical ensemble ( i.e. solution (II) ) 
is reduced to
\begin{eqnarray}
Z^{(II)}_{colorless}(\beta,V)&=&
\,
Z^{(0)}_{q\overline{q}g}(\beta,V) 
\,\times\,
\frac{1}{(2\pi)^{2}}\frac{1}{N!}
\,
\int^{\infty}_{-\infty} d\theta_{1}\, 
\int^{\infty}_{-\infty} d\theta_{2}
\,
\prod_{i<j}\left|\theta_{i}-\theta_{j}\right|^{2}
\nonumber\\
&~&~
\times
\exp\left[
-\frac{1}{2} a_{11}(\beta,V)\, \theta_{1}^{2}
-a_{12}(\beta,V)\, \theta_{1}\theta_{2}
-\frac{1}{2} a_{22}(\beta,V)\, \theta_{2}^{2}
\right],
\label{solution_nochiral_ii}
\end{eqnarray}
where
\begin{eqnarray}
Z^{(0)}_{q\overline{q}g}(\beta,V)&=&
\left.Z_{q\overline{q}g}(\beta,V;\theta_{1},\theta_{2})
\right|_{\theta_{1}=0,\theta_{2}=0},
\end{eqnarray}
and
\begin{eqnarray}
a_{11}(\beta,V)&=& 
\left.\frac{\partial^{2}}{\partial \theta_{1}^{2}}
\log_{e} Z_{q\overline{q}g}\left(\beta,V;\theta_{1},\theta_{2}\right)
\right|_{\theta_{1}=0,\theta_{2}=0},
\nonumber\\
a_{12}(\beta,V)&=&
\left.\frac{\partial^{2}}{\partial \theta_{1} \partial \theta_{2} }
\log_{e} Z_{q\overline{q}g}\left(\beta,V;\theta_{1},\theta_{2}\right)
\right|_{\theta_{1}=0,\theta_{2}=0},
\nonumber\\
a_{22}(\beta,V)&=&
\left.\frac{\partial^{2}}{ \partial \theta_{2}^{2} }
\log_{e} Z_{q\overline{q}g}\left(\beta,V;\theta_{1},\theta_{2}\right)
\right|_{\theta_{1}=0,\theta_{2}=0}.
\end{eqnarray}
Furthermore, in the case of massless flavors and zero flavor chemical potential 
(i.e. $\mu_{q}=0$), 
the canonical ensemble (II) for the high lying energy 
solution is simplified to
\begin{eqnarray}
Z^{(II)}_{colorless}\left(\beta,V\right)
&=&
\frac{\left(\prod^{N_{c}-1}_{n=1} n!\right)}
{\sqrt{N_{c}}(2\pi)^{\frac{N_{c}-1}{2}} }
\frac{
\exp\left[
\frac{V}{\beta^{3}}\left(
\frac{\pi^{2}}{45} \left(N^{2}_{c}-1\right)
+\frac{7\pi^{2}}{180} N_{c} N_{f}+\frac{1}{4\pi^{2}} 
N_{c} N_{f} \Lambda^{4}\beta^{4}
\right)\right]
}
{
\left[\frac{V}{\beta^{3}}
\left(
\frac{1}{3}N_{f}+\frac{2\pi^{2}}{45} \frac{{N_{c}}^{2}-1}{N_{c}}
\right)
\right]^{\frac{N^{2}_{c}}{2}-\frac{1}{2}}
}.
\label{solution-II-chiral-restored1}
\end{eqnarray}

In the present model, the order parameter(s) 
of GW-like phase transition 
is (are) temperature (and/or flavor chemical potentials).
At that critical point $T=T_{c}$, 
the low-lying and high-lying energy solutions match 
each other. 
In this case, the low-lying energy solution 
is extrapolated to the high-lying energy solution  
at the threshold of GW-like point.
The critical value of $T_{c}=1/\beta_{c}$ is determined 
by the continuity condition 
\begin{eqnarray}
\left.Z^{(I)}_{colorless}\left(\beta,V\right)\right|_{\beta=\beta_c}&=&
\left.Z^{(II)}_{colorless}\left(\beta,V\right)\right|_{\beta=\beta_c}.
\end{eqnarray}
Below GW-like point, the non-Gaussian stationary point 
of the asymptotic solution (I) is limited to $|\Phi|<1$.
At the threshold of GW-like phase transition, the 
non-Gaussian stationary points 
turn to be Gaussian saddle points that oscillate 
in the neighborhood of the center of the symmetry group.
Hence, the solution (II) turns to the asymptotic solution 
above GW-like point. 
The exact solution of Eq.(\ref{ex-colorless-projection})
is obtained by evaluating the integration over 
the invariance Haar measure numerically.
It is found that the asymptotic solution (I) 
matches the exact numerical solution below 
GW-like point while the solution (II) matches the exact 
one above GW-like point. Solutions (I) and (II) 
intersect each other in the neighborhood of GW-like point.
The Gaussian saddle points procedure is better understood 
in the terms of fundamental gauge field variables 
$\theta_{i}$ rather than Polyakov loop variables 
$\Phi$ and $\overline{\Phi}$. 
Furthermore, the deconfinement phase transition takes place 
when the high-lying energy states of quark-gluon bags 
become unstable and in this case the Hagedorn matter undergoes 
phase transition to quark-gluon plasma at Hagedorn's temperature. 
It is worth to note here that below Hagedorn's temperature, 
the high-lying energy quark-gluon bag
acts as quark-gluon fluid (or semi-QGP) droplets 
as far the constituent quarks and gluons remain within 
the range of the effective VanderMonde potential interaction.

\section{The extension to Polyakov-Nambu-Jona-Lasinio model\label{sectionIII}}
The conventional NJL Lagrangian density reads
\begin{eqnarray}
{\cal L}_{NJL}&=&\overline{q}\left[i\gamma^{\mu} \partial_{\mu}
-m_{q} \right] q + \frac{1}{2} G\left[
\left(\overline{q} q\right)^{2}+
\left(\overline{q} i\gamma_{5} \vec{\tau} q\right)^{2}
\right],
\end{eqnarray} 
where $m_{q}$ is the quark's current mass and 
$G$ is the NJL coupling constant. The constant $G$ is adjusted in 
order to fit the nuclear phenomenology.
The current mass and coupling constant 
for light flavors are taken 
$m_{q}=5$ MeV and $G = 10.992$, 
respectively.
The quark and antiquark grand potential density 
in the presence of the effective chiral field 
is furnished by  
\begin{eqnarray}
\frac{\Omega_{q\overline{q}}
\left(\beta,V;\sigma,\Phi,\overline{\Phi}\right)}{V}
&=&
-2 N_c
\sum^{N_f}_{q} 
\int \frac{d^{3} \vec{p}}{(2\pi)^{3}}
E_{q\sigma}(\vec{p})
-\frac{1}{\pi^2}
\sum^{N_{f}}_{q}
\int d|\vec{p}|
\, \frac{|\vec{p}|^{4}}{E_{q\sigma}(\vec{p})}
\nonumber
\\
&~& 
\times
\left(
\frac{
\left[
\left(
\Phi+2\overline{\Phi}
e^{-\beta\left[ E_{q\sigma}(\vec{p})-\mu_{q}\right]}
\right)         
e^{-\beta\left[ E_{q\sigma}(\vec{p})-\mu_{q}\right]}
+
e^{-3\beta\left[ E_{q\sigma}(\vec{p})-\mu_{q}\right]}
\right]
}
{
\left[
1+
3\left(\Phi+\overline{\Phi} 
e^{-\beta\left[ E_{q\sigma}(\vec{p})-\mu_{q}\right]}
\right)  
e^{-\beta\left[ E_{q\sigma}(\vec{p})-\mu_{q}\right]}
+  e^{-3\beta\left[ E_{q\sigma}(\vec{p})-\mu_{q}\right]}
\right]
}
\right.
\nonumber
\\
&~&
+
\left.
\frac{
\left[
\left(
\overline{\Phi}+2\Phi
e^{-\beta\left[ E_{q\sigma}(\vec{p})+\mu_{q}\right]}
\right)
e^{-\beta\left[ E_{q\sigma}(\vec{p})+\mu_{q}\right]}
+
e^{-3\beta\left[ E_{q\sigma}(\vec{p})+\mu_{q}\right]}
\right]
}
{
\left[1+
3\left(\overline{\Phi}+\Phi
e^{-\beta\left[ E_{q\sigma}(\vec{p})+\mu_{q}\right]}
\right)
e^{-\beta\left[ E_{q\sigma}(\vec{p})+\mu_{q}\right]}
+  
e^{-3\beta\left[ E_{q\sigma}(\vec{p})+\mu_{q}\right]}
\right]
}      
\right),
\label{PNJL-quark-antiquark-chiral-1}
\end{eqnarray}
where 
$E_{q\sigma}(\vec{p})=\sqrt{ \vec{p}^{2}+M_{q}(\sigma)^{2} }$ 
and $M_{q}(\sigma)=m_{q}-G\,\sigma$.
The parameters $m_{q}$, $\sigma$, $\Phi$, $G$ and $\mu_{q}$ are  
quark's current mass,  scalar field, Polyakov loop parameter, 
scalar coupling constant and 
constituent quark's chemical potential, respectively.
The $\sigma$ scalar mean field indicates 
the condensate $\sigma=<q\overline{q}>$.
The first term on the right hand side of 
Eq.(\ref{PNJL-quark-antiquark-chiral-1})
is temperature independent and 
is regularized as follows,
\begin{eqnarray}
\int^{\Lambda}_{0} \frac{d^{3}\vec{p}}{(2\pi)^{3}} 
E_{q\sigma}(\vec{p})
&=&
\frac{1}{2\pi^{2}}
\left[
\frac{1}{8} m_{q} \left(\sigma\right)^{2} \Lambda
\sqrt{ m_{q} \left(\sigma\right)^{2} + \Lambda^{2} }
+
\frac{1}{4} \Lambda^{3}
\sqrt{ m_{q} \left(\sigma\right)^{2} + \Lambda^{2} }
\right.
\nonumber \\
&~&
\left.
-\frac{1}{8} m_{q} \left(\sigma\right)^{4}
\log_{e}\left(
\sqrt{1+\frac{\Lambda^2}{m_{q} \left(\sigma\right)^2}}
+
\frac{\Lambda}{|m_{q} \left(\sigma\right)|}
\right)
\right],
\label{regularize-grand-term1}
\end{eqnarray}
where the UV-cutoff $\Lambda=631.5$ MeV. 
The scalar $\sigma$-chiral field in the context of PNJL model
is considered self-consistently.
In the terms of fundamental gauge fields 
$(\theta_{1},\theta_{2},\theta_{3}=-\theta_{1}-\theta_{2})$
rather than Polyakov loops ($\Phi$,$\overline{\Phi}$),
Eq.(\ref{PNJL-quark-antiquark-chiral-1})
is reduced to
\begin{eqnarray}
-\frac{\Omega_{q\overline{q}}\left(\beta,V;\sigma,\Phi,\overline{\Phi}\right)}{V}
&=&
-\frac{\Omega_{q\overline{q}}
\left(\beta,V;\sigma,
\Phi(\theta_{1},\theta_{2}),\overline{\Phi}(\theta_{1},\theta_{2})
\right)}{V},
\nonumber\\
&\rightarrow&
-\frac{\Omega_{q\overline{q}}
\left(\beta,V;\sigma,\theta_{1},\theta_{2}\right)}{V},
\end{eqnarray}
where
\begin{eqnarray}
&~&
-\frac{\Omega_{q\overline{q}}\left(\beta,V;\sigma,\theta_{1},\theta_{2}\right)}{V}
=
2 N_{c}\sum^{N_f}_{q} \int^{\Lambda}_{0} \frac{d|\vec{p}| |\vec{p}|^{2}}{2{\pi}^{2}}
E_{q\sigma}\left(\vec{p}\right)
\nonumber\\
&~&
~~~ ~~~ 
+ 2  \sum^{N_f}_{q} \sum^{N_{c}}_{k=1}
\int \frac{d |\vec{p}|}{(2\pi^{2})} \frac{|\vec{p}|^{4}}
{3 E_{q\sigma}\left(\vec{p}\right) }
\left(
\frac{ 1+ \cos(\theta_{k})\,
e^{\beta\left[E_{q\sigma}\left(\vec{p}\right)  -\mu_{q}\right]}
}{
1+ 2 \cos(\theta_{k})\,
e^{\beta\left[ E_{q\sigma}\left(\vec{p}\right)-\mu_{q}\right]}
+ e^{2\beta\left[E_{q\sigma}\left(\vec{p}\right)-\mu_{q}\right]}
}
\right.
\nonumber\\
&~& ~~~ ~~~ ~~~ ~~~ ~~~
+ \left.\frac{ 1+
\cos(\theta_{k})\,
e^{\beta\left[E_{q\sigma}\left(\vec{p}\right)+\mu_{q}\right]}
}{ 1+
2 \cos(\theta_{k})\,
e^{\beta\left[E_{q\sigma}\left(\vec{p}\right)+\mu_{q}\right]}
+
e^{2\beta\left[E_{q\sigma}\left(\vec{p}\right)+\mu_{q}\right]}
} \right)
\nonumber\\
&~& 
~~~ ~~~ 
+ \,i\, 2 \sum^{N_f}_{q} \sum^{N_{c}}_{k=1}
\int \frac{d |\vec{p}|}{(2\pi^{2})} \frac{|\vec{p}|^{4}}
{3 E_{q\sigma}\left(\vec{p}\right)}
\left( \frac{
e^{\beta\left[E_{q\sigma}\left(\vec{p}\right)-\mu_{q}\right]}
\sin(\theta_{k})
}{ 1+ 2 \cos(\theta_{k})\,
e^{\beta\left[E_{q\sigma}\left(\vec{p}\right)-\mu_{q}\right]}
+ e^{2\beta\left[E_{q\sigma}\left(\vec{p}\right)-\mu_{q}\right]}
} \right.
\nonumber\\
&~& ~~~ ~~~ ~~~ ~~~ ~~~
- \left. \frac{
e^{\beta\left[E_{q\sigma}\left(\vec{p}\right)+\mu_{q}\right]}
\sin\left(\theta_{k}\right)
}{ 1+ 2 \cos\left(\theta_{k}\right)\,
e^{\beta\left[E_{q\sigma}\left(\vec{p}\right)+\mu_{q}\right]}
+ e^{2\beta\left[E_{q\sigma}\left(\vec{p}\right)+\mu_{q}\right]}
} \right).
\label{quark-antiquark--eigen-ens2}
\end{eqnarray}
Therefore, in the case $\mu_{q}=0$, 
Eq.(\ref{quark-antiquark--eigen-ens2})
becomes real and is simplified to 
\begin{eqnarray}
&~&
-\frac{\Omega_{q\overline{q}}\left(\beta,V;\sigma,\theta_{1},\theta_{2}\right)}{V}
=
2 N_{c}\sum^{N_f}_{q} \int^{\Lambda}_{0} \frac{d|\vec{p}| |\vec{p}|^{2}}{2{\pi}^{2}}
E_{q\sigma}\left(\vec{p}\right)
\nonumber\\
&~&
~~~ ~~~ ~~~
+ 4  \sum^{N_f}_{q} \sum^{N_{c}}_{k=1}
\int \frac{d |\vec{p}|}{(2\pi^{2})} \frac{|\vec{p}|^{4}}
{3 E_{q\sigma}\left(\vec{p}\right) }
\left(
\frac{ 1+ \cos(\theta_{k})\,
e^{\beta\, E_{q\sigma}\left(\vec{p}\right)}
}{
1+ 2 \cos(\theta_{k})\,
e^{\beta\, E_{q\sigma}\left(\vec{p}\right)}
+ e^{2\beta\, E_{q\sigma}\left(\vec{p}\right)}
}
\right).
\label{quark-antiquark--eigen-ens-mu0}
\end{eqnarray}
Subsequently, the canonical ensemble for quarks and gluons 
in the context of the PNJL model becomes
\begin{eqnarray}
Z_{q\overline{q}g}\left(\beta,V;\sigma,\Phi,\overline{\Phi}\right)&=&
\exp\left[-\beta 
\Omega_{q\overline{q}g}\left(\beta,V;\sigma,\Phi,\overline{\Phi}\right)
\right].
\end{eqnarray}
The total grand potential for chiral quarks and gluons reads
\begin{eqnarray}
\Re e\,\Omega_{q\overline{q}g}\left(\beta,V;\sigma,\Phi,\overline{\Phi}\right)
&=&
\Omega_{q\overline{q}}\left(\beta,\sigma;\Phi,\overline{\Phi}\right)
+\Omega_{g}\left(\beta,V;\Phi,\overline{\Phi}\right)
+ V\,U\left(\sigma\right),
\end{eqnarray}
where $V$ is the system's volume.
The chiral quark and antiquark grand potential 
$\Omega_{q\overline{q}}(\beta,V;\sigma,\Phi,\overline{\Phi})$ 
is determined by
Eq.(\ref{PNJL-quark-antiquark-chiral-1}) 
while
the gluon grand potential 
$\Omega_{g}(\beta,V;\Phi,\overline{\Phi})$
is determined from
Eq.(\ref{phen-gluon-partition1}).
The effective chiral potential is given by
\begin{eqnarray}
U\left(\sigma\right)=
\frac{1}{2} G\, \sigma^{2}.
\label{effective-chiral-pot1}
\end{eqnarray}

The canonical ensemble for the colorless quark and gluon blob 
is determined by projecting 
the color-singlet state in the following way,
\begin{eqnarray}
Z_{colorless}\left(\beta,V;\sigma\right)&=&
\int d\mu\left({\bf g}\right)\, 
Z_{q\overline{q}g}\left(\beta,V;\sigma,\Phi,\overline{\Phi}\right),
\nonumber\\
&=&
\int d\mu\left({\bf g}\right)\,
Z_{q\overline{q}g}\left(\beta,V;\sigma,\theta_{1},\theta_{2}\right).
\label{colorless-chiral-qqg}
\end{eqnarray}
It is possible to write Eq.(\ref{colorless-chiral-qqg})
in the terms of Polyakov loop variables 
$\Phi$ and $\overline{\Phi}$ as follows
\begin{eqnarray}
Z_{colorless}\left(\beta,V;\sigma\right)&=&
N_{\mbox{Haar}}
\int_{C} d\Phi\,\int_{C} d\overline{\Phi}\, 
Z_{PNJL}\left(\beta,V;\sigma,\Phi,\overline{\Phi}\right),
\label{colorless-Polya-haar-1}
\end{eqnarray}
where the subscript $C$ indicates the integration over 
the three pointed star boundary in the complex plane 
and
\begin{eqnarray}
Z_{PNJL}
\left(\beta,V;\sigma,\Phi,\overline{\Phi}\right)
&=&
\exp\left(-\beta\left[
{\cal V}_{PVdM}\left(\beta;\Phi,\overline{\Phi}\right)+
\Omega_{q\overline{q}g}
\left(\beta,V;\sigma,\Phi,\overline{\Phi}\right)
\right] 
\right).
\label{low-lying-grand-PNJL-1}
\end{eqnarray}
The double integrations over Polyakov loop variables, namely, 
$\Phi,\overline{\Phi}$ is evaluated using 
the non-Gaussian stationary points method 
for the low-lying energy limit.
Subsequently using the non-Gaussian stationary 
points method,
Eq.(\ref{colorless-Polya-haar-1}) is reduced to
the low-lying energy solution as follows,
\begin{eqnarray}
Z^{(I)}_{colorless}\left(\beta,V;\sigma\right)&=&
Z^{(I)}_{colorless}
\left(\beta,V;\sigma,\Phi_{0},\overline{\Phi}_{0}\right),
\nonumber\\
&=&
Z_{PNJL}\left(\beta,V;\sigma,\Phi_{0},\overline{\Phi}_{0}\right),
\nonumber\\
&=&
\exp\left(-\beta\left[
{\cal V}_{PVdM}\left(\beta;{\Phi_{0}},{\overline{\Phi}_{0}}\right)
+\Omega_{q\overline{q}g}
\left(\beta,V;\sigma,{\Phi_{0}},{\overline{\Phi}_{0}}\right)
\right]
\right).
\label{low-lying-grand-PNJL-Phi1}
\end{eqnarray}
The pre-factor $N_{\mbox{Haar}}$ is eliminated 
in order to guarantee the normalization of 
the non-Gaussian stationary point method 
and it does not affect the calculation. 
The Polyakov loops' non-Gaussian stationary points, 
namely,
$\Phi_{0}$ and $\overline{\Phi}_{0}$ 
are determined by extremizing the exponent 
which appears on  the right hand side 
of Eq.(\ref{low-lying-grand-PNJL-1})
with respect to $\Phi$ 
and $\overline{\Phi}$ in the following way,
\begin{eqnarray}
\begin{array}{c}
\left.
\frac{\partial}{\partial \Phi}
\left[
{\cal V}_{PVdM}\left(\beta;\Phi,\overline{\Phi}\right)
+\Omega_{q\overline{q}g}
\left(\beta,V;\sigma,\Phi,\overline{\Phi}\right)
\right]
\right|_{\Phi=\Phi_{0},\overline{\Phi}=\overline{\Phi}_{0}}=0,
\\
\left.
\frac{\partial}{\partial \overline{\Phi}}
\left[
{\cal V}_{PVdM}\left(\beta;\Phi,\overline{\Phi}\right)
+\Omega_{q\overline{q}g}
\left(\beta,V;\sigma,\Phi,\overline{\Phi}\right)
\right]\right|_{\Phi=\Phi_{0},\overline{\Phi}=\overline{\Phi}_{0}}=0.
\end{array}
\label{PNJL-Phi-Phi-bar-extrema-1}
\end{eqnarray}
Furthermore, the $\sigma$-chiral field stationary point, 
namely $\sigma_{0}$, 
is determined by extremizing the exponent which
appears on the right hand side of 
Eq.(\ref{low-lying-grand-PNJL-Phi1}) 
with respect to the scalar field $\sigma$. 
Since Polyakov VanderMonde potential, namely, 
${\cal V}_{PVdM}\left(\beta;\Phi,\overline{\Phi}\right)$ 
does not depend on $\sigma$,
the variation of 
$\Omega_{q\overline{q}g}
\left(\beta,V;\sigma,\Phi_{0},\overline{\Phi}_{0}\right)$
with respect to $\sigma$ mean field leads to
\begin{eqnarray}
\sigma=
-\frac{1}{G}
\left.\frac{\partial}{\partial\sigma} 
\left(
\frac{
\Omega_{q\overline{q}}
\left(\beta,V;\sigma,\Phi_{0},\overline{\Phi}_{0}\right)
}{V}
\right)\right|_{\sigma=\sigma_{0}},
\label{PNJL-sigma-extrema-cal-1}
\end{eqnarray}
where
\begin{eqnarray}
&~&
- \frac{\partial}{\partial\sigma} 
\left(  \frac{ \Omega_{q\overline{q}}
\left(\beta,V;\sigma,\Phi,\overline{\Phi}\right)
}{V}  \right) = \sum^{N_{f}}_{q} \left(
M_{q}(\sigma) 
\frac{\partial}{\partial \sigma}  M_{q}(\sigma)\right)
\left[ 
2 N_{c}\int^{\Lambda}_{0} 
\frac{ d|\vec{p}| }{ 2 \pi^{2} }
\frac{ |\vec{p}|^{2} }{ E_{q\sigma}(\vec{p}) }
\right.
\nonumber\\
&~&
~~~~ ~~~~ ~~~~ ~~~~ ~~~~
-6 \int^{\infty}_{0} 
\frac{ d |\vec{p}| }{2\pi^{2}}
\frac{ |\vec{p}|^{2} }{ E_{q\sigma}(\vec{p}) }
\frac{
\left(\Phi+2\overline{\Phi} e^{ -\beta E_{q\sigma}(\vec{p}) }\right) 
e^{ -\beta E_{q\sigma}(\vec{p}) }+e^{ -3\beta E_{q\sigma}(\vec{p}) }
}{
[1+3\left(\Phi+\overline{\Phi} e^{ -\beta E_{q\sigma}(\vec{p}) }\right) 
e^{ -\beta E_{q\sigma}(\vec{p}) }+e^{ -3\beta E_{q\sigma}(\vec{p}) }
]}
\nonumber\\
&~&
~~~~ ~~~~ ~~~~ ~~~~ ~~~~
\left.
-
6 \int^{\infty}_{0} 
\frac{d|\vec{p}|}{2\pi^{2}}
\frac{ {|\vec{p}|}^{2} }{ E_{q\sigma}(\vec{p}) }  
\frac{
\left( \overline{\Phi} + 2 \Phi e^{-\beta E_{q\sigma}(\vec{p}) }\right) 
e^{ -\beta E_{q\sigma}(\vec{p}) }+e^{-3\beta E_{q\sigma}(\vec{p}) }
}{[1+3\left( \overline{\Phi}  + \Phi e^{-\beta E_{q\sigma}(\vec{p}) }\right) 
e^{ -\beta E_{q\sigma}(\vec{p}) }+e^{-3\beta E_{q\sigma}(\vec{p}) }
]}
\right].
\label{PNJL-grand-sig-extrema}
\end{eqnarray}
Moreover, when Eq.(\ref{PNJL-grand-sig-extrema})
is written in the terms of fundamental gauge fields 
$(\theta_{1},\theta_{2})$, it is reduced to
\begin{eqnarray}
&~&
- \frac{\partial}{\partial\sigma} 
\left(  \frac{ \Omega_{q\overline{q}}
\left(\beta,V;\sigma,\theta_{1},\theta_{2}\right)}{V}  \right) 
=  \sum^{N_{f}}_{q} \left( M_{q}(\sigma) 
\frac{\partial}{\partial \sigma}  M_{q}(\sigma)\right)
\left[ 
2 N_{c}\int^{\Lambda}_{0} 
\frac{ d|\vec{p}| }{ 2 \pi^{2} }
\frac{ |\vec{p}|^{2} }{ E_{q\sigma}(\vec{p}) }
\right.
\nonumber\\
&~&
~~~~ ~~~~ ~~~~ ~~~~ ~~~~
\left.
-4\,\sum^{N_{c}}_{k=1}\,\int^{\infty}_{0} 
\frac{ d |\vec{p}| }{2\pi^{2}}
\frac{ |\vec{p}|^{2} }{ E_{q\sigma}(\vec{p}) }
\,\frac{1+e^{\beta\, E_{q\sigma}(\vec{p})}\cos\left(\theta_{k}\right)
}{
1+2e^{\beta\,E_{q\sigma}(\vec{p})}\cos\left(\theta_{k}\right)
+e^{2\beta\,E_{q\sigma}(\vec{p}) }
}
\right].
\label{Haar-grand-sig-extrema}
\end{eqnarray}
The first term inside the square bracket on the right hand side 
of Eq.(\ref{PNJL-grand-sig-extrema}) 
and/or Eq.(\ref{Haar-grand-sig-extrema})
is temperature independent. 
Its explicit expression reads 
\begin{eqnarray}
\int^{\Lambda}_{0} \frac{d |\vec{p}| }{2\pi^{2}}
\frac{ {|\vec{p}|}^{2} }{ E_{q\sigma}(\vec{p}) }
&=&
\frac{1}{4\pi^{2}}
\Lambda  \sqrt{M_{q}(\sigma)^2+\Lambda^2}
\nonumber\\
&-&
\frac{1}{4\pi^{2}}
M_{q}(\sigma)^2
\log_{e}\left(
\frac{\Lambda}{M_{q}(\sigma)}
+
\sqrt{
1+\left(\frac{\Lambda}{M_{q}(\sigma)}\right)^{2}
}
\right).
\end{eqnarray}
Therefore, the partition function (i.e. the canonical ensemble) 
for the colorless quark and gluon bag 
with $\sigma$-chiral field reads
\begin{eqnarray}
Z^{(I)}_{colorless}\left(\beta,V\right)&=& 
Z^{(I)}_{colorless}\left(\beta,V;\sigma_{0}\right),
\nonumber\\
&=&
Z^{(I)}_{colorless}
\left(\beta,V;\sigma_{0},\Phi_{0},\overline{\Phi}_{0}\right),
\nonumber\\
&=&
Z_{PNJL}
\left(\beta,V;\sigma_{0},\Phi_{0},\overline{\Phi}_{0}\right),
\end{eqnarray}
where the values of $\Phi_{0}$, $\overline{\Phi}_{0}$ 
and $\sigma_{0}$ are the stationary points 
and they are calculated by 
Eqs.(\ref{PNJL-Phi-Phi-bar-extrema-1})
and (\ref{PNJL-sigma-extrema-cal-1}), respectively.
Evidently, 
when the temperature approaches the critical one 
(i.e. GW-like point),
the non-Gaussian stationary point method 
fails due to the logarithmic divergence
of Polyakov VanderMonde potential, namely,
${\cal V}_{PVdM}\left(\beta;\Phi,\overline{\Phi}\right)$.
The logarithmic divergence of Polyakov VanderMonde
potential indicates a collapse 
of the non-Gaussian stationary point method
for the low-lying energy solution and 
the emergence of the high-lying energy solution
where Polyakov loops' stationary points switch 
to become Gaussian saddle points that oscillate harmonically 
around the non-Gaussian stationary points. 
The Gaussian saddle point procedure is understood 
in the context of fundamental gauge fields 
$\theta_{1}$ and $\theta_{2}$
much better than in the frame work of Polyakov loops 
$\Phi$ and $\overline{\Phi}$.

On the other hand, the partition function 
for the asymptotic high-lying energy solution 
(i.e. solution II)
for colorless quark and gluon blob 
in the context of PNJL model reads
\begin{eqnarray}
Z^{(II)}_{colorless}\left(\beta,V;\sigma\right)&=&
\int d\mu\left({\bf g}\right)\, 
Z_{q\overline{q}g}\left(\beta,V;\sigma,\Phi,\overline{\Phi}\right),
\nonumber\\
&=&
\frac{1}{(2\pi)^{2}}\frac{1}{N!}
\int^{\infty}_{-\infty} d\theta_{1}\int^{\infty}_{-\infty}  d\theta_{2}
\,\prod_{i<j}\left(\theta_{i}-\theta_{j}\right)^{2}
\,e^{
-\beta\, \Omega_{q\overline{q}g}
\left(\beta,V;\sigma,\theta_{1},\theta_{2}\right)
},
\label{colorless-high-lying-integral-1}
\end{eqnarray}
where $\theta_{3}=-\theta_{1}-\theta_{2}$. 
The quark and gluon grand potential 
which appears in the exponent in 
Eq.(\ref{colorless-high-lying-integral-1}) 
is given by
\begin{eqnarray}
\Omega_{q\overline{q}g}
\left(\beta,V;\sigma,\theta_{1},\theta_{2})\right)&=&
\Omega_{q\overline{q}g}
\left(\beta,V;\sigma,
\Phi(\theta_{1},\theta_{2}),\overline{\Phi}(\theta_{1},\theta_{2})
\right),
\nonumber\\
&=& 
\Omega_{q\overline{q}}\left(\beta,V;\sigma,\theta_{1},\theta_{2}\right)
+
\Omega_{g}\left(\beta,V;\sigma,\theta_{1},\theta_{2}\right)
+V\,U(\sigma),
\end{eqnarray}
where $V$ is the bag's volume 
and Polyakov loop parameters 
$\Phi$ and $\overline{\Phi}$ are written explicitly as functions of
the fundamental gauge fields $\theta_{1}$, $\theta_{2}$ and $\theta_{3}$. 
The effective scalar potential, namely $U(\sigma)$, 
is given by Eq.(\ref{effective-chiral-pot1}).
Since the non-Gaussian stationary points for the low-lying 
energy solution (I) are reduced to Gaussian saddle points 
for the asymptotic high-lying energy solution (II),
it becomes essential to compute the quadratic expansion 
of the grand potential around the Gaussian saddle points
in order to evaluate the integral that is given 
in Eq.(\ref{colorless-high-lying-integral-1}) 
more appropriately. 
This can be done much easier 
in the framework of fundamental gauge fields 
rather than Polyakov loops.
The Gaussian saddle points of the fundamental gauge 
fields accumulate at the origin and 
fortunately this behavior simplifies 
the calculation drastically.
The quadratic Taylor expansion of the grand potential density 
for quarks and anti-quarks is reduced to
\begin{eqnarray}
\frac{1}{V}
\Omega_{q\overline{q}}\left(\beta,V;\sigma,\theta_{1},\theta_{2}\right)
&=&
\frac{1}{V}
\Omega^{(0)}_{q\overline{q}}\left(\beta,V;\sigma\right)
+\frac{1}{2}
\frac{1}{V}
\Omega^{(2)}_{q\overline{q}}\left(\beta,V;\sigma\right)
\sum^{N_c}_{i}\theta_{i}^{2},
\end{eqnarray}
where the $0^{\mbox{th}}$ term reads
\begin{eqnarray}
\frac{1}{V}\Omega^{(0)}_{q\overline{q}}\left(\beta,V;\sigma\right)&=&
- 2 N_c\,\sum^{N_f}_{q}
\int^{\Lambda}_{0} \frac{d^{3}\vec{p}}{(2\pi)^{3}} 
\,
E_{q\sigma}(\vec{p})
\nonumber\\
&~&
- 4 N_{c} \sum^{N_f}_{q}
\int \frac{d|\vec{p}|}{ 2\pi^{2} }\,
\frac{|\vec{p}|^{4}}{ 3 E_{q\sigma}(\vec{p}) }\,
\frac{1}{\left[e^{ \beta E_{q\sigma}(\vec{p}) }+1\right]},
\end{eqnarray}
while the quadratic term is given by
\begin{eqnarray}
\frac{1}{V}\Omega^{(2)}_{q\overline{q}}\left(\beta,V;\sigma\right)
&=&
4\frac{1}{\beta} \sum^{N_f}_{q}
\int \frac{d|\vec{p}|}{ 2\pi^{2} }\,
|\vec{p}|^{2}\,
\frac{e^{ \beta E_{q\sigma}(\vec{p}) }}{
\left( e^{ \beta E_{q\sigma}(\vec{p}) }+1 \right)^{2}
}.
\end{eqnarray}
Again, the gluonic grand potential 
for the low-lying energy colorless quark-gluon bags 
is assumed to be adjusted by the phenomenology
as done in Sec.\ref{sectionII}
(see for instance Eq.(\ref{phen-gluon-partition1})).
This class of the phenomenological gluon potential 
is inspired 
from lattice calculations and has been recently 
adopted widely in the literature 
(for instance see~\cite{Ratti:2005jh}).
The quadratic Taylor expansion of the gluon grand potential
which is given by Eq.(\ref{phen-gluon-partition1})
is approximated to
\begin{eqnarray}
\frac{1}{V}
\Omega_{g}\left(\beta,V\right)&=&
\frac{1}{V}
\Omega^{(0)}_{g}\left(\beta,V\right)
+\frac{1}{2} 
\frac{1}{V}
\Omega^{(2)}_{g}\left(\beta,V\right)
\sum_{ij}\left(\theta_{i}-\theta_{j}\right)^{2}.
\end{eqnarray}
The $0^{\mbox{th}}$ term reads
\begin{eqnarray}
\frac{1}{V}\Omega^{(0)}_{g}\left(\beta,V\right)&=&
-\frac{1}{\beta}\left(\frac{\omega_{g}}{\beta^{3}}\right),
\end{eqnarray}
while the quadratic term is reduced to
\begin{eqnarray}
\frac{1}{V}\Omega^{(2)}_{g}\left(\beta,V\right)&=&
\frac{1}{\beta}\,\frac{1}{N^{2}_{c}}\,
\left(\frac{\omega_{g}}{\beta^{3}}\right),
\end{eqnarray}
where $\omega_{g}=\frac{\pi^{2} \left(N^{2}_{c}-1\right)}{45}$.  
%
After evaluating the Gaussian integration over 
the fundamental gauge fields, the partition function 
for the high-lying energy solution 
(i.e. solution II) is approximated to
\begin{eqnarray}
Z^{(II)}_{colorless}\left(\beta,V\right)
&=& Z^{(II)}_{colorless}\left(\beta,V;\sigma_{0}\right),
\end{eqnarray}
where
\begin{eqnarray}
Z^{(II)}_{colorless}\left(\beta,V;\sigma\right)
&=&
\frac{
\left(\prod^{N_{c}-1}_{n=1} n!\right) 
}
{ \sqrt{N_{c}} (2\pi)^{\frac{1}{2}\left(N_{c}-1\right)} } 
\frac{
\exp\left(-\beta\left[
\Omega^{(0)}_{q\overline{q}}\left(\beta,V;\sigma\right)
+\Omega^{(0)}_{g}\left(\beta,V\right)
+V\,U(\sigma)
\right]\right)}
{
\left(
\beta\left[
\Omega^{(2)}_{q\overline{q}}\left(\beta,V;\sigma\right)
+2 N_{c}\Omega^{(2)}_{g}\left(\beta,V\right)
\right]
\right)^{\frac{N_{c}^{2}-1}{2}}}.
\label{PNJL-II}
\end{eqnarray}
Furthermore, 
$\sigma$-mean field
(i.e. $\sigma_{0}$), above 
the threshold of GW-like phase transition, 
is determined by calculating 
$\sigma$-stationary point in the following way, 
\begin{eqnarray}
\left.\frac{\partial}{\partial \sigma} 
\frac{
\Omega^{(II)*}_{q\overline{q}g}\left(\beta,V;\sigma\right)
}{V}
\right|_{\sigma=\sigma_{0}}
&=&0,
\label{PNJL-II-sigma-0-1}
\end{eqnarray}
where 
\begin{eqnarray}
\Omega^{(II)*}_{q\overline{q}g}\left(\beta,V;\sigma\right) 
&=&
\Omega^{(0)}_{q\overline{q}}(\beta,V;\sigma)
+\Omega^{(0)}_{g}(\beta,V)
+V\,U(\sigma).
\label{grand-potential-II-1}
\end{eqnarray}
Under the assumption of the stationary point method, 
the extremization procedure is performed 
for the exponent term that appears in 
Eq.(\ref{PNJL-II}).
The extremization of Eq.(\ref{PNJL-II-sigma-0-1})
leads to
\begin{eqnarray}
G\sigma&=&
-\frac{\partial}{\partial \sigma}
\left(\frac{
\Omega^{(0)}_{q\overline{q}}\left(\beta,V;\sigma\right)
}{V}
\right),
\label{chiral-PNJL-sigma-ii}
\end{eqnarray}
where
\begin{eqnarray}
-\frac{\partial}{\partial \sigma}
\left(
\frac{
\Omega^{(0)}_{q\overline{q}}\left(\beta,V;\sigma\right)
}{V}
\right)
&=&
2 N_c
\sum^{N_f}_{q}
\left[
\int^{\Lambda}_{0} \frac{d|\vec{p}|}{2\pi^{2}}
\frac{ {\vec{p}}^{2} }{ E_{q\sigma}(\vec{p}) }
-2 \int \frac{d|\vec{p}|}{2\pi^{2}}
\frac{ {\vec{p}}^{2} }{ E_{q\sigma}(\vec{p}) }
\frac{1}{\left(e^{ \beta E_{q\sigma}(\vec{p}) }+1\right)}
\right]
\nonumber\\
&~&
~~~ ~~~ \times
\left(M_{q}(\sigma)
\frac{\partial M_{q}(\sigma)}{\partial\sigma}
\right).
\end{eqnarray}
In order to calculate other thermodynamics quantities,
the derivative of the partition function 
with respect to $X$ is reduced to 
\begin{eqnarray}
&-&\frac{\partial}{\partial X}\frac{1}{\beta} 
\log_{e}\left[
Z^{(II)}_{colorless}\left(\beta,V;\sigma\right)\right]
=
\frac{\partial}{\partial X} \Omega^{(0)}_{q\overline{q}}(\beta,V;\sigma)
+\frac{\partial}{\partial X} \Omega^{(0)}_{g}(\beta,V)
+ \frac{\partial}{\partial X}\left[V\,U(\sigma) \right]
\nonumber\\
&~&
~~~ ~~~ ~~~ ~~~
+\left(\alpha-\frac{1}{2}\right) \frac{\partial}{\partial X}
\left\{
\frac{1}{\beta} 
\log_{e}\left(\beta\left[
\Omega^{(2)}_{q\overline{q}}(\beta,V;\sigma)      
+ 2 N_{c} \Omega^{(2)}_{g}(\beta,V)
\right]\right)
\right\},
\label{grand-canonical-potential-1}
\end{eqnarray}
where $X$ is a thermodynamic ensemble such as $\beta$ and $V$.
For instance, from Eq.(\ref{grand-canonical-potential-1}), 
the grand potential for colorless quark and gluon blob 
reads
\begin{eqnarray}
\frac{1}{V}\Omega^{(II)}_{q\overline{q}g}\left(\beta,V;\sigma\right)
&=& - \frac{\partial}{\partial V}\frac{1}{\beta}
\log_{e}\left[
Z^{(II)}_{colorless} \left(\beta,V;\sigma\right) \right],
\nonumber\\
&=& 
\frac{ \Omega^{(0)}_{q\overline{q}}(\beta,V;\sigma) }{V}
+\frac{ \Omega^{(0)}_{g}(\beta,V) }{V}
+U(\sigma) - \frac{ \alpha-\frac{1}{2} }{V}.
\label{grand-potential-II-0}
\end{eqnarray}
Hence, if the $\sigma$-chiral field is not restored below GW-like point, 
then it will be a discontinuity 
(i.e. at least of a higher order discontinuity) 
from $\sigma_{0}=\sigma^{(I)}_{0}$ 
to $\sigma_{0}=\sigma^{(II)}_{0}$
in the neighborhood of GW-like point because
the value of $\sigma^{(I)}_{0}$ below GW-like point 
is determined by
Eq.(\ref{PNJL-sigma-extrema-cal-1})
while $\sigma^{(II)}_{0}$ above GW-like point 
is determined 
by Eq.(\ref{PNJL-II-sigma-0-1}) 
or Eq.(\ref{chiral-PNJL-sigma-ii}).
However, the extensive numerical calculations 
show that the chiral symmetry restoration 
usually occurs in the neighborhood 
of GW-like point and before the ultimate point 
of the extended GW-range is reached.
This indicates that the high-lying energy solution 
is chirally restored. 
It should be noted that in infinite volume limit, the VanderMonde
regularization becomes essential and, subsequently, 
Eq.(\ref{PNJL-II}) is reduced to
\begin{eqnarray}
Z^{(II)}_{colorless}\left(\beta,V\right)
&=& Z^{(II)}_{colorless}\left(\beta,V;\sigma_{0}\right),
\end{eqnarray}
where
\begin{eqnarray}
Z^{(II)}_{colorless}\left(\beta,V;\sigma\right)
&=&
\frac{
\left(\prod^{N_{c}-1}_{n=1} n!\right)
}
{ \sqrt{N_{c}} (2\pi)^{\frac{1}{2}\left(N_{c}-1\right)} }
\frac{
\exp\left(-\beta\,\Omega^{(II)*}_{q\overline{q}g}\left(\beta,V;\sigma\right)
\right)}
{
\left(
\beta\left[
\Omega^{(2)}_{q\overline{q}}\left(\beta,V;\sigma\right)
+2 N_{c}\Omega^{(2)}_{g}\left(\beta,V\right)
\right]
\right)^{ \gamma_{reg}\frac{(N_{c}^{2}-1)}{2} }},
\label{PNJL-II-regularized}
\end{eqnarray}
where $\gamma_{reg}=\frac{V}{a^{3}}$ 
and $a^{3}$ is the lattice space size. 
Nonetheless, the regularization procedure 
is not required for
finite colorless quark and gluon bag.

The order parameter for GW-like phase transition is 
the temperature (and the chemical potentials). 
The point of the phase transition, namely, $T_{GW}$
is determined by the continuity of the partition function
from the low-lying energy solution to the high-lying one 
and this condition is satisfied when both solutions 
match each others as follows, 
\begin{eqnarray}
\begin{array}{c}
Z^{(I)}_{colorless}\left(\beta,V\right)
=
\left. Z^{(II)}_{colorless}\left(\beta,V\right) 
\right|_{\beta=\beta_{GW}}
\\
\rightarrow
Z^{(I)}_{colorless}\left(\beta_{GW},V;\sigma^{(I)}\right)
=
Z^{(II)}_{colorless}\left(\beta_{GW},V;\sigma^{(II)}\right).
\end{array}
\end{eqnarray}
The values of $\sigma^{(I)}$ and $\sigma^{(II)}$
are determined using Eqs.(\ref{PNJL-sigma-extrema-cal-1}) 
and (\ref{chiral-PNJL-sigma-ii}), respectively.
The both solutions (I) and (II) are asymptotic solutions 
for the low and high temperatures, respectively. 
This implies that the solution (I)'s partition function 
is extrapolated to the solution (II) 
when the threshold of GW-like point is reached.
Beyond that point, the solution (I) 
deviates significantly from the exact numerical one 
and turns to be no longer correct. 
Fortunately, the solution (II) provides a clue whereabouts 
GW-like point threshold and its ultimate point
and also their interpolation range 
(i.e. the interval between the threshold and ultimate point). 
The Helmholtz free energy of solution (II) has 
a hidden valley. 
The validity of the asymptotic solution (II) is maintained
whenever the energy climbs the hidden valley and reaches 
the same level of its virtual top that appears 
at lower temperature. 
This point is ultimate point of the extended interval 
of GW-like point. 
Beyond the ultimate point, solution (II) matches
the exact one precisely.
When solution (II) intersects solution (I), 
the threshold of an extended GW-like interval emerges. 
At GW-threshold, the solution (I) starts to deviate significantly
and subsequently solution (II) turns to be the correct asymptotic 
solution instead of solution (I). 
The deviation becomes significant when GW-ultimate point is reached.
Therefore, it is reasonable to interpolate solution (I) 
from GW-threshold to the asymptotic solution (II) 
at GW-ultimate point.

The PNJL-partition function can be solved exactly. 
The partition function for the colorless quark and gluon blob reads
\begin{eqnarray}
Z_{colorless}\left(\beta,V\right)&=& 
Z_{colorless}\left(\beta,V;\sigma_{0}\right),
\end{eqnarray}
where
\begin{eqnarray}
Z_{colorless}\left(\beta,V;\sigma\right)
&=&
\int^{\pi}_{-\pi} d\theta_{1}\,\int^{\pi}_{-\pi} d\theta_{2}
\,\mu_{\mbox{Haar}}\left(\theta_{1},\theta_{2}\right)
\, \exp\left[ -\beta\,
\Omega_{q\overline{q}g}
\left(\beta,V;\sigma,\theta_{1},\theta_{2}\right)
\right].
\end{eqnarray}
The invariance Haar measure is given by
\begin{eqnarray}
\mu_{\mbox{Haar}}\left(\theta_{1},\theta_{2}\right)&=&
\frac{\prod^{(N_{c}-1)}_{n=1} n!}{N_{c}!(2\pi)^{N_{c}-1}}
\prod_{i<j}
4 \sin\left(\frac{\theta_{i}-\theta_{j}}{2}\right)^2.
\end{eqnarray}
Furthermore, chiral mean field, namely, 
$\sigma_{0}$ is evaluated by extremizing  
the partition function as follows
\begin{eqnarray}
-\left.
\frac{\partial}{\partial \sigma}
\frac{1}{V \beta}
\log_{e} 
\left[
Z_{colorless}\left(\beta,V;\sigma\right)
\right]
\right|_{\sigma=\sigma_{0}}&=& 0.
\label{exact-PNJL-sigma-sol}
\end{eqnarray}
Thus Eq.(\ref{exact-PNJL-sigma-sol}) 
is reduced to
\begin{eqnarray}
&~&
\int^{\pi}_{-\pi} d\theta_{1}\,\int^{\pi}_{-\pi} d\theta_{2}
\,\mu_{\mbox{Haar}}\left(\theta_{1},\theta_{2}\right)
\,\exp\left[ -\beta\,
\Omega_{q\overline{q}g}
\left(\beta,V;\sigma,\theta_{1},\theta_{2}\right)
\right]
\nonumber\\
&~& ~~~ ~~~\times
\left[
\frac{\partial}{\partial \sigma}
\frac{\Omega_{q\overline{q}g}
\left(\beta,V;\sigma,\theta_{1},\theta_{2}\right)}{V}
\right]=0.
\label{exact-PNJL-sigma-sol2}
\end{eqnarray}
Hence by using
Eqs.(\ref{exact-PNJL-sigma-sol}) and (\ref{exact-PNJL-sigma-sol2})
the $\sigma_{0}$-chiral mean field is determined 
by solving the following equation,
\begin{eqnarray}
\frac{\partial}{\partial\sigma} U(\sigma)
&=& 
\frac{1}{
Z_{colorless}\left(\beta,V;\sigma\right)
}
\,
\int^{\pi}_{-\pi} d\theta_{1}\,\int^{\pi}_{-\pi} d\theta_{2}
\,\mu_{\mbox{Haar}}\left(\theta_{1},\theta_{2}\right)
\nonumber\\
&~& ~\times
\left[
-
\frac{\partial}{\partial \sigma} 
\frac{\Omega_{q\overline{q}}\left(\beta,V;\sigma,\theta_{1},\theta_{2}\right)}{V}
\right]
\,\exp\left[ -\beta\,
\Omega_{q\overline{q}g}\left(\beta,V;\sigma,\theta_{1},\theta_{2}\right)
\right].
\label{exact-PNJL-sigma-sol-eq}
\end{eqnarray}
\section{GW-like point and Hagedorn states}

The asymptotic mass spectral density of states 
is given by the micro-canonical ensemble. 
The micro-canonical ensemble can be derived 
from the mixed-grand canonical ensemble 
of a single QG-bag.
It is given by the inverse Laplace transform 
as follows
\begin{eqnarray}
\rho_{colorless}\left(W,V\right)&\sim&
\frac{1}{2\pi\, i} 
\int^{\beta_{0}+i\,\infty}_{\beta_{0}-i\,\infty}
\, d\beta\,
e^{\beta\, W}\, Z_{colorless}\left(\beta,V\right),
\label{micro-canonical-ensemble-1}
\end{eqnarray}
where $W$ is the energy of QG-bag.
In the limit of large $W$, 
Eq.(\ref{micro-canonical-ensemble-1}) 
is evaluated using the steepest descent method. 
The approximation of the steepest descent method fails
in the limit of small $W$. 
This means that it is reasonable to replace 
the low-lying mass spectral density
with the discrete mass spectrum 
of the hadron states while the high-lying 
mass spectral density in the large $W$ limit 
is approximated to the bootstrap-like mass spectral 
density for Hagedorn states. 
Hence, it is more appropriate to replace solution (I) 
with the discrete mass spectrum of hadronic states.
Furthermore, it will be shown below that the extrapolation 
of the mass spectral density of solution (I) 
to Hagedorn states 
does not lead to a deconfinement phase transition 
to QGP at Hagedorn's temperature. 
In contrary, the mass spectral density for solution (II) 
leads to a first order phase transition.  
Therefore, the existence of Hagedorn states 
is interpreted in the term of GW-like phase transition
where the discrete hadronic mass spectrum turns 
to the continuous bootstrap-like mass spectrum 
when the hadron's mass exceeds 
a specific mass threshold (i.e. $m_{H}>2$ GeV). 
In order to simplify the calculation drastically, 
the chiral field is dropped in this section. 
The density of states for solution (I) 
is reduced to 
\begin{eqnarray}
\lim_{W\rightarrow \infty}
\rho^{(I)}_{colorless}\left(W,V\right)
&\sim&
\frac{1}{2\pi\, i}
\int^{\beta_{0}+i\,\infty}_{\beta_{0}-i\,\infty}\, d\beta\,
e^{\beta\, W}\, Z^{(I)}_{colorless}\left(\beta,V\right),
\nonumber\\
&\sim&
\frac{1}{2\pi\, i}
\int^{\beta_{0}+i\,\infty}_{\beta_{0}-i\,\infty}\, d\beta\,
e^{\beta\, W}\, 
e^{-\beta {\cal V}_{PVdM}\left(\beta;\Phi_{0},\overline{\Phi}_{0}\right)
-\beta \Omega_{q\overline{q}g}
\left(\beta,V;\Phi_{0},\overline{\Phi}_{0}\right)},
\nonumber\\
&\sim&
\frac{
e^{ W_{PVdM}\left(\Phi_{0},\overline{\Phi}_{0}\right) }
}
{2\pi\, i}
\int^{\beta_{0}+i\,\infty}_{\beta_{0}-i\,\infty}\, d\beta\,
e^{\beta\, W}\,
e^{\frac{V}{\beta^{3}} a_{q\overline{q}g} 
\left(\Phi_{0},\overline{\Phi}_{0}\right)
+
\frac{N_{c} N_{f} \Lambda^{4}}{4\pi^{2}}V\beta},
\nonumber\\
&\sim&
\frac{
e^{ W_{PVdM}\left(\Phi_{0},\overline{\Phi}_{0}\right) }
}
{2\pi\, i}
\int^{\beta_{0}+i\,\infty}_{\beta_{0}-i\,\infty}\, d\beta\,
e^{\beta\, W'}\,
e^{\frac{V}{\beta^{3}} a_{q\overline{q}g} 
\left(\Phi_{0},\overline{\Phi}_{0}\right)
},
\label{micro-I-canonical-ensemble-1}
\end{eqnarray}
where 
$W'\,=\,W+\frac{N_{c} N_{f} \Lambda^{4}}{4\pi^{2}}V$
and
\begin{eqnarray}
W_{PVdM}\left(\Phi_{0},\overline{\Phi}_{0}\right)
&=&
\frac{1}{2}\log_{e}\left(
1-6\Phi_{0} \overline{\Phi}_{0} + 
4\left(\Phi_{0}^{3}+\overline{\Phi}_{0}^{3}\right)
-3 \left(\Phi_{0}\overline{\Phi}_{0}\right)^{2}
\right),
\end{eqnarray}
and  
\begin{eqnarray}
a_{q\overline{q}g}
\left(\Phi_{0},\overline{\Phi}_{0}\right)
&=&
2 N_{f}\int^{\infty}_{0} \frac{d x\, x^{3}}{2\pi^{2}}
\left[
\frac{\left(\Phi_{0}+2\overline{\Phi}_{0} e^{-x}\right)e^{-x}+e^{-3x}}
{
1+\left(\Phi_{0}+\overline{\Phi}_{0} e^{-x}\right)e^{-x}+e^{-3x}
}
\right.
\nonumber\\
&~& ~~~
\left.
+
\frac{\left(\overline{\Phi}_{0}+ 2\Phi_{0} e^{-x}\right)e^{-x}+e^{-3x}}
{ 
1+\left(\overline{\Phi}_{0}+\Phi_{0} e^{-x}\right)e^{-x}+e^{-3x}
}
\right]
+\omega_{g} \Phi_{0} \overline{\Phi}_{0}.
\label{a_qqg_phi_1}
\end{eqnarray}
In the limit of $\Phi_{0},\overline{\Phi}_{0}\rightarrow 0$,
Eq.(\ref{a_qqg_phi_1}) is simplified to
\begin{eqnarray}
\lim_{\Phi_{0},\overline{\Phi}_{0}\rightarrow 0}
a_{q\overline{q}g} \left(\Phi_{0},\overline{\Phi}_{0}\right)
&=&
\frac{N_{f}}{81} \, \left(\frac{7 \pi^{2}}{60}\right).
\end{eqnarray}
Under the assumption of MIT bag model and 
in the limit of $\Phi_{0},\overline{\Phi}_{0}\rightarrow 0$,
the extrapolation of the mass spectral density (I) 
is reduced to 
\begin{eqnarray}
\rho^{(I)}_{colorless}\left(m\right)
&\sim&
C\, { \beta_{(I)} }^{5/2}
\,m^{-1/2}\,e^{ b \, m },
\end{eqnarray}
where 
$m=W'\,+\,BV$ 
and $B^{\frac{1}{4}}\sim 200-250$ MeV is the bag constant 
and
\begin{eqnarray}
\beta_{(I)}&=&
\left(
\frac{N_{f}}{3^{3}} \, \frac{7 \pi^{2}}{60}
\frac{1}{3 B}
\right)^{1/4},
\nonumber\\
b&=& \beta_{(I)},
\nonumber\\
C&=&
\frac{1}{2\sqrt{2\pi}} 
\left(
\frac{4B}
{
\frac{N_{f}}{3^{3}} \, \frac{7 \pi^{2}}{60}
}
\right)^{1/2}.
\end{eqnarray}
It is more appropriate to represent the large $W$ limit 
in the term of asymptotic solution (II).
Under the assumption of solution (II), 
Hagedorn's density of states is approximated to
\begin{eqnarray}
\lim_{W\rightarrow \infty} 
\rho^{(II)}_{colorless}\left(W,V\right)&\sim&
\frac{1}{2\pi\, i}
\int^{\beta_{0}+i\,\infty}_{\beta_{0}-i\,\infty}\, d\beta\,
e^{\beta\, W}\, Z^{(II)}_{colorless}\left(\beta,V\right).
\label{micro-II-canonical-ensemble-1}
\end{eqnarray}
In the context of MIT bag model, 
Eq.(\ref{micro-II-canonical-ensemble-1}) 
is reduced to
\begin{eqnarray}
\rho^{(II)}_{colorless}\left(m\right)&\sim&
C\,
{\beta_{(II)}}^{\frac{3}{2}N_{c}^{2}+1}
\, m^{-N_{c}^{2}/2}\,
e^{b\, m},
\label{micro-II-canonical-ensemble-2}
\end{eqnarray}
where
\begin{eqnarray}
\beta_{(II)}&=&
\left(
\frac{
\frac{\pi^{2}}{15} \left(N^{2}_{c}-1\right)
+\frac{7\pi^{2}}{60} N_{c} N_{f}
}
{3 B}
\right)^{1/4},
\nonumber\\
b&=& \beta_{(II)},
\nonumber\\
C&=&
\frac{1}{2\sqrt{2\pi}}
\, \left(4 B\right)^{\frac{N_{c}^{2}}{2}} \,
\frac{\left(\prod^{N_{c}-1}_{n=1} n!\right)}{\sqrt{N_{c}}(2\pi)^{\frac{N_{c}-1}{2}} }
\, \frac{  \left(
\frac{1}{3}N_{f}+\frac{2\pi^{2}}{45} \frac{{N_{c}}^{2}-1}{N_{c}}
\right)^{-\frac{ N^{2}_{c}}{2}+\frac{1}{2} }  }
{\left(
\frac{\pi^{2}}{15} \left(N^{2}_{c}-1\right)
+\frac{7\pi^{2}}{60} N_{c} N_{f}\right)^{\frac{1}{2}}}.
\end{eqnarray}
It is interesting to note that the mass spectral density 
for solution (I) does not lead to a deconfinement phase 
transition at Hagedorn's temperature 
while the system with mass spectral density (II)
undergoes a first order deconfinement phase transition.
For a system with two flavors (i.e. $N_f=2$) 
and $B^{1/4}=250$ MeV, 
Hagedorn's temperature for the deconfinement 
phase transition is reduced to $T_{H}\sim 608$
MeV and $176$ MeV for solutions (I) and (II), respectively.
With smaller bag constant $B^{1/4}=200$ MeV, 
Hagedorn's temperature is reduced to $T_{H}\sim 487$
MeV and $141$ MeV for solutions (I) and (II), respectively.
Hagedorn's temperature for the solution (II) 
is more acceptable than that for solution (I).
When the exponent $\alpha$ in 
$\rho(m)\propto m^{-\alpha} e^{b\, m}$
runs over $5/2<\alpha\le 7/2$, 
Hagedorn matter undergoes a higher order phase transition
while the system undergoes a first order phase transition 
for $7/2<\alpha$.
Therefore, GW-like phase transition is interpreted 
as an extrapolation of the discrete mass spectrum 
of the conventional hadronic states that are found 
in the data book ~\cite{databook:2008} to Hagedorn states 
(i.e. super massive hadronic states) 
that are represented by the bootstrap-like models.
In this context, the deconfinement phase transition 
to QGP takes place at Hagedorn's temperature. 
In this sense, GW-like transition is a hadronic mechanism
that produces meta-stable super-massive hadronic states 
(known as Hagedorn states) 
below the deconfinement phase transition to QGP. 
Finally, it should be noted that the regularization 
procedure for VanderMonde's number of states
reduces the spectral density to
\begin{eqnarray}
\rho^{(II)}_{colorless}\left(m\right)
&\propto&
\, m^{-\alpha}\, e^{b\, m},
\nonumber\\
&\propto&
\, m^{
-\left(\frac{\gamma_{reg}}{2}\,(N_{c}^{2}-1)+\frac{1}{2}
\right)
}\,
e^{b\, m}.
\label{micro-II-canonical-regularized}
\end{eqnarray}
Eq.(\ref{micro-II-canonical-regularized}) 
demonstrates that $\gamma_{reg}$ 
may be related to the bag's volume fluctuation.
It is reduced to $\gamma_{reg}=1$ 
for a bag with a sharp surface boundary. 
The cases $\gamma_{reg}<1$ and $\gamma_{reg}>1$
correspond to the expanding (dilute) and squeezing 
(compressed) bags, respectively. 
The case $\gamma_{reg}<1$ is related to the bag 
with an extended surface boundary. 
The exponent $\alpha$ is reduced to
$\frac{9}{2}$ and $\frac{3}{2}$
for $\gamma_{reg}=1$ and 
$\gamma_{reg}=\frac{1}{4}$, 
respectively.  

\section{Discussion and conclusion}

We have considered the canonical ensemble for colorless quark and gluon blob.
The colorless quark and gluon blobs emerge as meta-stable Hagedorn states
in the relativistic heavy ion collisions. 
These colorless states (i.e. Hagedorn states) significantly 
enrich the deconfinement phase transition diagram.
Their production signature may mix and be confused with QGP.
In order to make the discussion simple, 
at first we neglect the effect of chiral field and simply
assume massless 2-flavors in order to simplify the analysis 
of GW-like phase transition. 
The low-lying energy solution, namely, solution (I),
is determined by non-Gaussian stationary point method 
for Polyakov loop parameters ($\Phi$, $\overline{\Phi}$) 
as defined by Eq.(\ref{ensemble-Polyakov-Haar-stationary1}).
The high-lying energy solution, namely, solution (II), 
is determined by the Gaussian saddle point approximation.
The assumption is that the non-Gaussian stationary points 
of solution (I) turn to Gaussian saddle points in solution (II). 
The solution (II) is introduced by 
Eq.(\ref{solution_nochiral_ii}).
Furthermore, the exact numerical solution is considered by evaluating
the exact numerical integration over the fundamental gauge fields 
$\theta_{1}$ and $\theta_{2}$ with the invariance Haar measure 
which is given by Eq.(\ref{ensemble-Polyakov-Haar1}).

\begin{figure}
\includegraphics{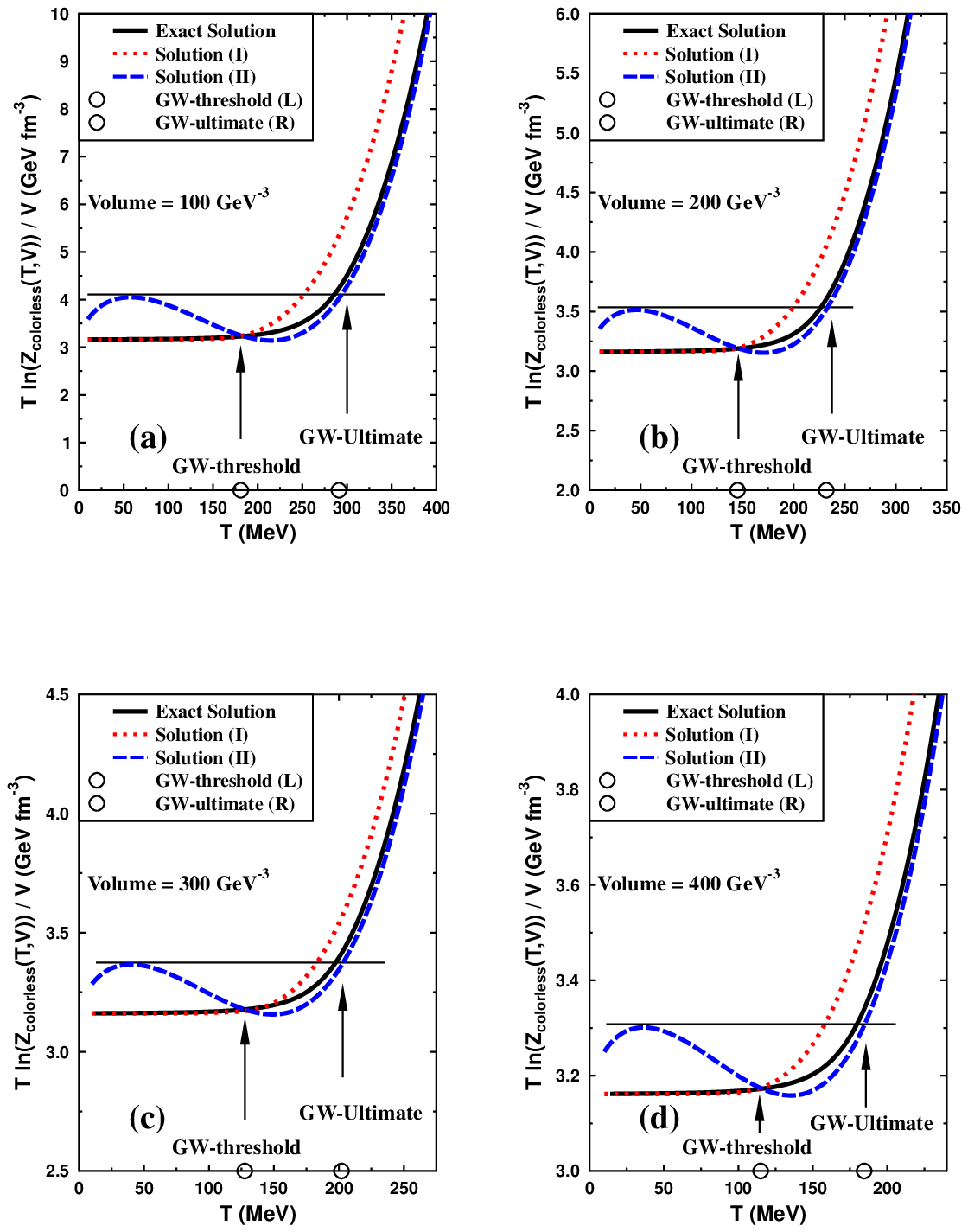}
\caption{ 
\label{GW-point-nochiral}
(Color online)
The {\em negative} Helmholtz free energy density 
$\frac{T}{V}\log_{e} Z_{colorless}\left(T,V\right)$
for the low-lying energy (I) and high-lying energy (II) solutions
and the exact one 
(i.e. the phase transition solution (I) $\rightarrow$ (II) between
GW-threshold and ultimate points) 
vs temperature. No chiral field is considered.
The GW-threshold and ultimate points 
are shown as circles in the $T$-axis.
The results for volumes 100, 200, 300 and 400 GeV$^{-3}$
(i.e. $R=$ 0.57, 0.71, 0.82 and 0.90 fm)
are displayed respectively in (a), (b), (c) and (d), respectively.}
\end{figure}

Fig.~\ref{GW-point-nochiral} depicts the quantity
$\frac{T}{V} \log_{e} Z_{colorless}\left(T,V\right)$ 
which represents the {\em negative} Helmholtz free energy density
vs $T$ for quark and gluon blob with various volumes 
$R=0.57$, $R=0.71$, $R=0.82$ and $R=0.90$ fm.
It is shown that solution (I) matches
exact numerical solution below GW-threshold temperature 
and then it deviates from the exact one when the 
temperature reaches and exceeds GW-threshold point
while solution (II) converges to the exact numerical solution 
as temperature approaches GW-ultimate point until 
it fits precisely the exact one as temperature exceeds 
that point.
Therefore, solution (I) is the correct asymptotic solution 
for any temperature below GW-threshold point 
while solution (II) is the correct asymptotic 
solution for any temperatures above GW-ultimate point.
Furthermore, it seems that neither solution (I) nor solution (II) 
fits correctly the exact numerical solution over 
an extended GW-like point domain
which covers the interval between GW-threshold and ultimate points. 
Evidently, the interpolation of both solutions (I) and (II) 
over the interval between GW-threshold and ultimate points
fits the exact numerical solution. 
This makes a smooth transition from solution (I) 
to solution (II)
over an extended GW-like point interval.
The domain between threshold and ultimate points 
(i.e. over the extended GW-like point interval)
is reduced to a single point in the limit 
$N_{c}\rightarrow \infty$ but a finite coupling constant 
$g\, N^{2}_{c}$.
Therefore, the analytical solution is modified {\em smoothly} 
from solution (I) to solution (II) 
over the extended GW-point interval.
This implies that low-lying and high-lying mass spectra 
remain in mutual exchange reaction
over the extended GW-point interval.
The high-lying energy solution
(II) significantly deviates from the exact solution 
at temperature below GW-threshold and then turns 
to converge to the exact one as the temperature 
approaches GW-ultimate point and then remains 
in an excellent match 
as the temperature increases beyond GW-ultimate point. 
On the other hand, the low-lying energy solution 
(I) matches the exact numerical solution precisely 
for temperature below GW-threshold 
and then it starts to deviate significantly 
from the exact one
when the temperature exceeds GW-threshold.
This deviation is significant 
as temperature increases above GW-ultimate point.
The {\em smooth} modification 
in the solution's analytic behavior through
the extended GW-point interval clearly implies 
that Hagedorn states emerge as meta-stable states 
over an extended GW-point interval 
with mutual and exchange chemical reaction between
the high-lying and low-lying hadronic states.
The exchange reaction clarifies the difficulty 
to detect Hagedorn states, 
GW-like transition and the subsequent confusion 
with the deconfinement phase transition. 

\begin{figure}
\includegraphics{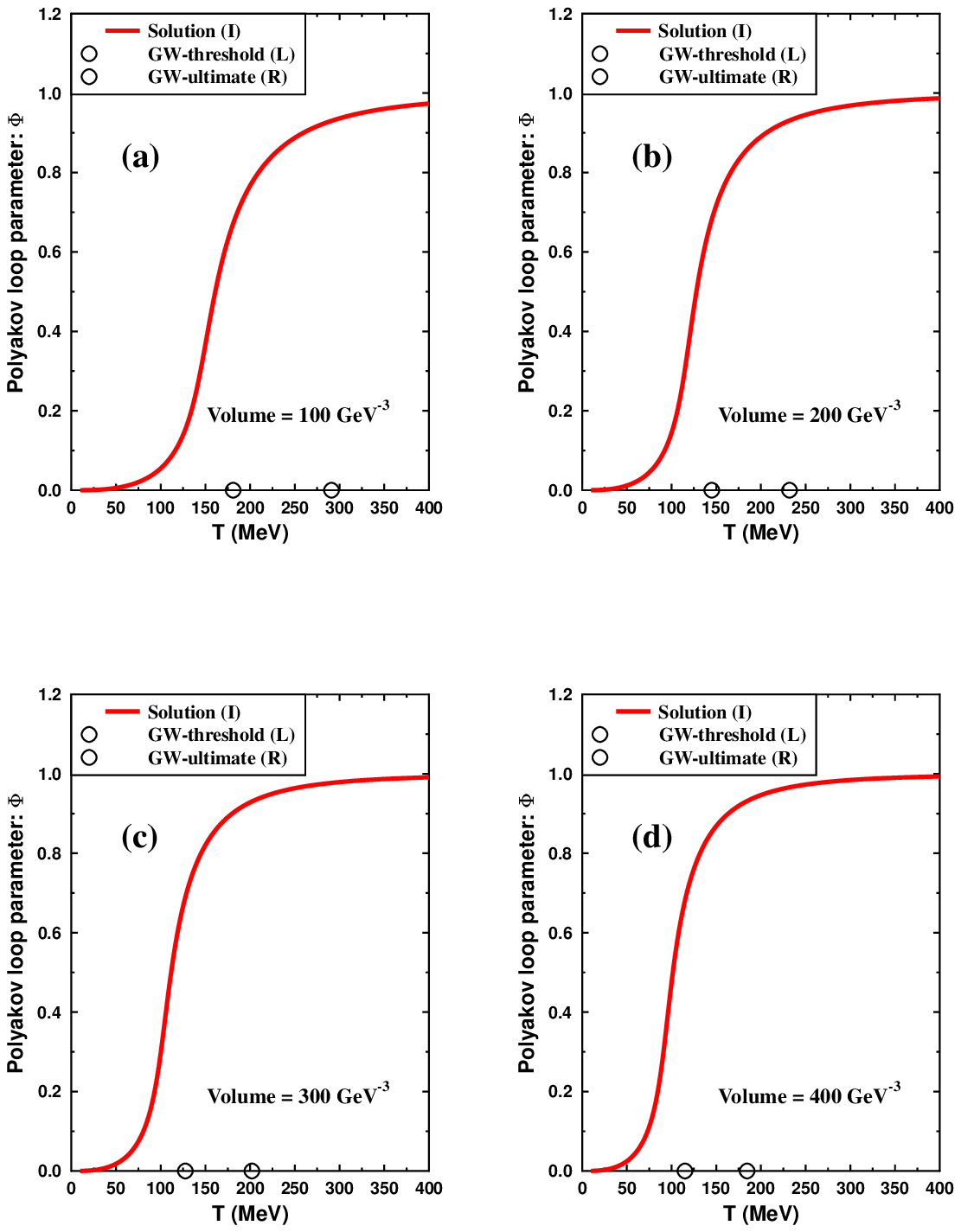}
\caption{
\label{Polyakovloop-wchiral}
(Color online)
The Polyakov loop (triality) parameter $\Phi$, the order parameter of solution (I), 
vs temperature with various bag's volume. The chiral field is not included.
(a) $V = 100 \mbox{GeV}^{-3}$. (b) $V = 200 \mbox{GeV}^{-3}$.
(c) $V = 300 \mbox{GeV}^{-3}$. (d) $V = 400 \mbox{GeV}^{-3}$.}
\end{figure}

\begin{figure}
\includegraphics{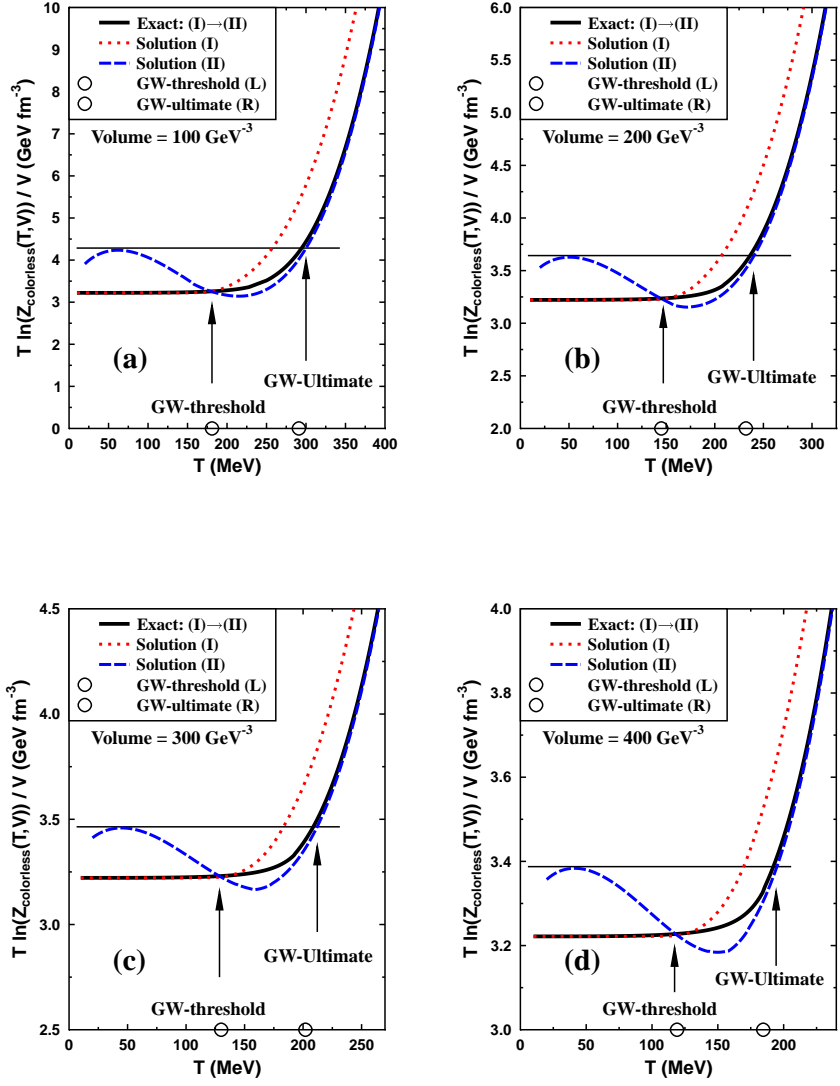}
\caption{
\label{Polyakov-Z-wchiral}
(Color online)
The {\em negative} Helmholtz free energy density 
$\frac{T}{V}\log_{e} Z_{colorless}\left(T,V\right)$
for the low-lying energy (I) and high-lying energy (II) solutions
and exact one vs temperature.
The $\sigma$-chiral mean field is included simultaneously 
in the calculation.
The GW-threshold and ultimate points are shown 
as circles in the $T$-axis.
(a) $V = 100 \mbox{GeV}^{-3}$ $(R =0.57 \mbox{fm})$. 
(b) $V = 200 \mbox{GeV}^{-3}$ $(R =0.71 \mbox{fm})$.
(c) $V = 300 \mbox{GeV}^{-3}$  $(R =0.82 \mbox{fm})$. 
(d) $V = 400 \mbox{GeV}^{-3}$ $(R =0.90 \mbox{fm})$.}
\end{figure}

\begin{figure}
\includegraphics{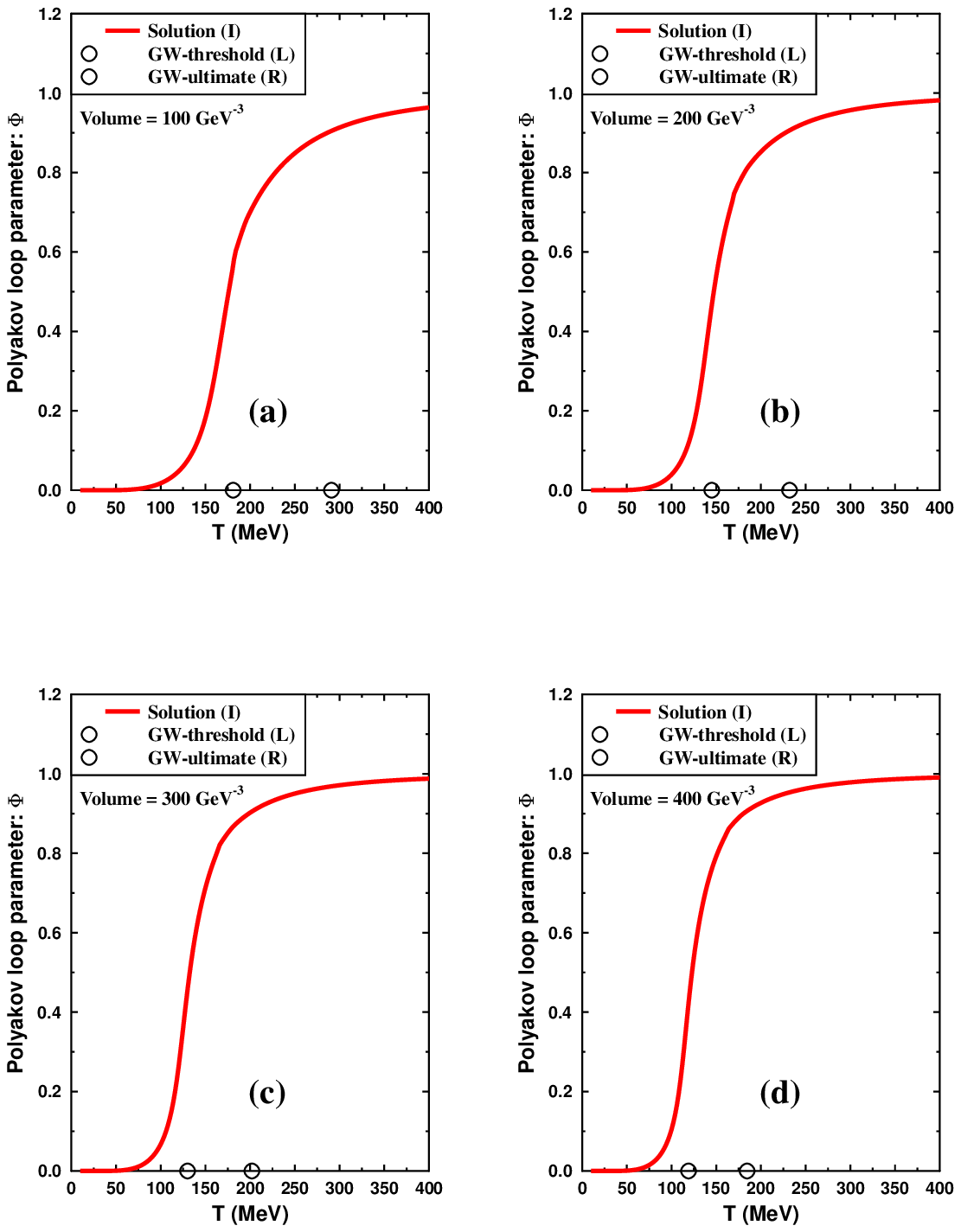}
\caption{
\label{Polyakov-triality-wchiral}
(Color online)
The Polyakov loop (triality) parameter $\Phi$,
the order parameter of solution (I),
vs temperature with various bag's volume.
The $\sigma$-chiral mean field 
is included simultaneously in the calculation.
(a) $V = 100 \mbox{GeV}^{-3}$. (b) $V = 200 \mbox{GeV}^{-3}$.
(c) $V = 300 \mbox{GeV}^{-3}$. (d) $V = 400 \mbox{GeV}^{-3}$.}
\end{figure}

The order parameter $\Phi_{0}$ for solution (I) 
vs temperature is depicted 
in Fig.~\ref{Polyakovloop-wchiral}
with various volumes of colorless quark and gluon bags.
The order parameter $\Phi_{0}$ is simply  
the stationary point that projects the color singlet state 
under the assumption of solution (I).
The order parameter $\Phi_{0}$ is found very small  
at low temperatures 
and this is because of the strong confinement.
This implies a reduced gluonic component 
for hadronic states at low temperatures since 
$\Phi_{0}$ and $\overline{\Phi}_{0}$ correspond 
the gluon condensates. 
Furthermore, when the system is heated up, 
the value of $\Phi_{0}$ increases and approaches 
its restoration value 
from below but remains $\Phi_{0}<1$.  
The $\Phi_{0}$'s asymptotic restoration
indicates loose confinement states or meta-stable bubbles. 
Furthermore, $\Phi_{0}$ increases 
from $\Phi_{0}\approx 0^{+}$
to the restoration value $\Phi_{0}\approx 1^{-}$ 
within the extended GW-point interval 
as the asymptotic solution switches from (I) to (II).
Although, GW-like phase transition from strong coupling to weak coupling 
has been extensively considered in the context of large-$N_{c}$ limit,
it is evident that GW-like phase transition 
persists to exist even in QCD with $N_{c}=3$ but 
with different analytical behavior. 
The GW-like phase transition in QCD is not a conventional 
confinement/deconfinement phase transition 
but is the Hagedorn's production threshold.  
This can be understood in the term of micro-canonical ensemble
and the consideration of gas of Hagedorn states.
The micro-canonical ensemble of solution (II) 
is the mass spectral density of Hagedorn states 
where GW-like point corresponds the Hagedorn's mass threshold 
(i.e. $m_{threshold}\sim$ 2 GeV).
Therefore, the present results
suggest that GW-like phase transition persists to exist
in QCD and, subsequently, the Polyakov loop restoration 
turns to be the onset of GW-like phase transition 
or semi-QGP and the existence of Hagedorn states where 
Hagedorn states are produced in the hadronic phase.
This interpretation, definitely, implies that 
the Polyakov loop restoration is not 
the deconfinement's order parameter 
as has been suggested in some models~\cite{Fukushima:2003fw} 
but the abundant production of (meta-stable) Hagedorn states
below Hagedorn's temperature.
This conclusion is also true for PNJL model where 
the $\sigma$-chiral field 
is considered explicitly and self-consistently in the calculation.
Fig.\ref{Polyakov-Z-wchiral} displays 
the {\em negative} Helmholtz free energy 
vs temperature with various volumes. 
The general situation looks very similar to Fig.~\ref{GW-point-nochiral}. 
The solution (I) matches the exact numerical one 
for temperature below GW-threshold.  
When the temperature exceeds GW-threshold point, solution (I)
starts to deviate significantly above the exact one. 
It continues to deviate above the exact numerical one 
as the temperature increases.
On the other hand, the high-lying energy solution, namely solution (II), 
has a hidden valley that deviates significantly 
from the exact numerical one at low temperature
as far the temperature remains below GW-ultimate limit. 
This valley emerges due to the unphysical oscillatory 
behavior of the Gaussian saddle point approximation below GW-like point.
Nonetheless, solutions (I) and (II) intersect with each others 
at GW-threshold temperature below GW-ultimate temperature. 
When the temperature increases and reaches GW-ultimate point, 
solution (II) converges to and matches precisely 
the exact numerical one.
Furthermore, as the temperature increases and exceeds 
GW-ultimate point, 
solution (II) converges to exact numerical one 
and remains in an excellent agreement. 
Therefore, evidently there is a switch  
from solution (I) to solution (II) within the extended 
GW-point interval.
Nevertheless, the extended GW-point interval is ambiguous 
in heavy ion collisions 
as far neither solution (I) nor solution (II) fits the exact one 
while their interpolation seems to fit to the exact numerical solution.  
The importance of this mechanism is that it may shed the light on 
the existence of (meta-) Hagedorn states and their production 
as super-massive hadronic states (i.e. $m_{H}>2$ GeV).
The extended GW-point interval 
(i.e. the interval between the threshold and ultimate points) 
is very sensitive to the fireball's volume.
For instance, the extended GW-point interval 
is extended from $T_{min}$ to $T_{max}$ ($\sim 181 - 291$ MeV) 
for bag's radius $R =0.57$ fm.
The extended GW-point interval is significantly 
reduced and turns to $\sim 119-184.5$ MeV 
for bag's radius $R =0.90$ fm. 
Hence, Hagedorn states turn 
to be of the size of QGP 
(i.e. $R \ge 0.90$ fm)
for temperature close to the deconfinement 
one (i.e. $T\approx 184.5$ MeV).
Furthermore, Hagedorn states with size $R =0.57$ fm 
are likely to be produced at rather high temperature 
$T_{max}=291$ MeV while large Hagedorn states 
are produced at lower temperatures.
This unusual behavior makes more difficult 
to detect Hagedorn states
as far they emerge as super-massive, gluonic rich
and meta-stable states with the size order of QGP.
The large (volume and mass) Hagedorn states 
can be confused and mixed with a true deconfinement 
phase transition's signature.
The (super-)massive Hagedorn states can be developed 
as droplets of semi-quark-gluon plasma in colorless states.
Fig.~\ref{Polyakov-triality-wchiral} depicts 
the order parameter $\Phi_{0}$ for solution (I)
vs temperature with various  volumes.
The restoration of Polyakov loop $\Phi_{0}$
likely takes place over the extended GW-point 
interval 
(i.e. between GW-threshold and ultimate points).
When $T$ approaches GW-threshold, $\Phi_{0}$ 
(and $\overline{\Phi}_{0}$) 
starts its significant restoration process. 
Furthermore, when $T$ exceeds GW-ultimate point
$\Phi_{0}$ turns to be almost restored from below 
(i.e. $\Phi_{0}\le 1$).
This behavior hints that the non-Gaussian stationary 
point approximation fails at temperature above GW-threshold.
Subsequently, 
the non-Gaussian stationary point approximation
is converted to Gaussian saddle point approximation.

\begin{figure}
\includegraphics{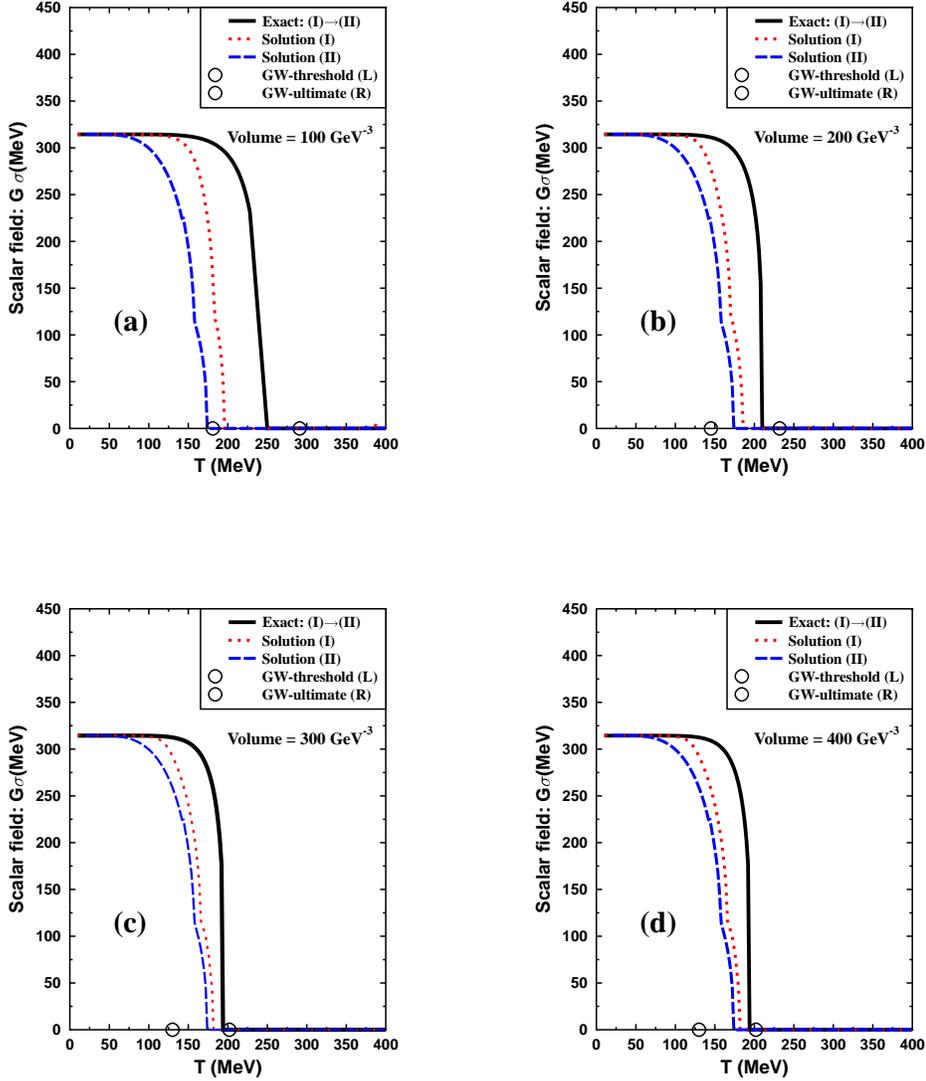}
\caption{
\label{scalarfield-wchiral}
(Color online)
The chiral mass $G\,\sigma$ vs temperature for the low-lying (I) and high-lying (II)
energy solutions and the exact one  
(i.e. the phase transition solution (I) $\rightarrow$ (II) between
GW-threshold and ultimate points) 
with various bag's volume.
The $\sigma$-chiral mean field and 
Polyakov loops are included simultaneously in the calculation.
(a) $V = 100 \mbox{GeV}^{-3}$. (b) $V = 200 \mbox{GeV}^{-3}$.
(c) $V = 300 \mbox{GeV}^{-3}$. (d) $V = 400 \mbox{GeV}^{-3}$.}
\end{figure}

The effective chiral field $G\,\sigma $ vs $T$ 
with various volumes is displayed 
in Fig.~\ref{scalarfield-wchiral}.
The $\sigma$-chiral mean field is considered 
self-consistently in the frame work of 
solutions (I) and (II) as well as the exact numerical solution. 
The results show clearly that the chiral restoration
likely takes place within an extend GW-point interval 
but below GW-ultimate point. 
Furthermore, the solution (II)'s chiral restoration 
takes place before that one for solution (I).
This implies that 
chiral restoration likely takes place 
within an extended GW-point interval 
above GW-threshold point but below GW-ultimate point. 
The both solutions (I) and (II) fail to locate 
the precise position of chiral restoration. 
This deficit is understood by realizing that 
extrapolation of solution (I) or (II) 
is not the correct asymptotic solution 
over the extended GW-point interval. 
Furthermore, the chiral restoration of the exact numerical solution 
usually occurs on the right hand side of solutions (I) and (II) 
but below GW-ultimate point.  
This can be interpreted as exchange reaction
and smooth transition between the low-lying and high-lying 
hadronic states over the extended GW-point interval. 
The chiral restoration takes place within an extended 
GW-point interval but far away from GW-ultimate point 
for small fireball ($R\sim 0.57$ fm). 
The restoration point approaches GW-ultimate point 
from below as the fireball size increases.
Hence, Hagedorn threshold production takes place 
within an extended GW-point interval.
The results demonstrate that the (meta-) Hagedorn 
states are chirally restored above GW-ultimate.

The results suggest that there is 
a new class of phase transition in nuclear matter 
in particular in the hadronic sector. 
The Hadronic phase turns to be smoothly dominated by 
a gas of colorless quark-gluon bags 
through the extended GW-point interval. 
This mechanism can be understood 
in the term of Hagedorn states. 
The size of Hagedorn's bags continue to grow up  
until Hagedorn's temperature is reached.
When Hagedorn's temperature is reached, the system 
undergoes a deconfinement phase transition to QGP.
The finite volume colorless states have been suggested 
before by Elze, Greiner and Rafelski 
~\cite{Elze:1986db,Elze:1986gz,Elze:1983du,Elze:1984un,
Elze:1985wv}. 
This picture has been extended to the gas of bags. 
The results also suggest a possible production 
of large (meta-) colorless quark-gluon droplets
of the size order of quark-gluon plasma $R\ge 0.90$ fm
at $T\le 184.5$ MeV (in the case of a single droplet analysis). 
The smaller Hagedorn states are produced 
at much higher temperatures. 
For instance bags with $R\sim 0.57$ fm
are produced at $T\sim 291$ MeV.
This supports that small size hadronic states 
belong to the low-lying hadronic mass spectrum 
rather than 
high-lying hadronic mass spectrum.  
In the case of gas of bags, the analysis can 
be extended using Hagedorn's density of states
that is derived from the micro-canonical ensemble.
This indicates that the nuclear matter 
undergoes smooth transition from low-lying mass spectrum 
to (meta-) Hagedorn states (or even semi-QGP) 
rather than directly to true deconfined QGP. 
The deconfinement phase transition takes place 
at Hagedorn's temperature.
The production of colorless QG-fireballs enriches 
the nuclear phase transition diagram significantly. 
This mechanism opens a window to produce (meta-)stable 
colorless super-massive QG-droplets at the size of order of QGP.
Finally, the signature of deconfined QGP 
in the heavy ion collisions may be confused and/or mixed 
with the gas of colorless QG-bags or semi-QGP. 
\begin{acknowledgments}
This work was supported by Helmholtz International Centre for FAIR 
within the
framework of the LOEWE program (Landesoffensive zur
Entwicklung Wissenschaftlich-\"Okonomischer Exzellenz)
launched by the State of Hesse is acknowledged. 
One of us (IZ) thanks C. Sasaki, R. Pisraski and E. Witten
for the discussion. 

\end{acknowledgments}

\bibliography{refbib1}


\newpage

\newpage

\newpage

\newpage

\newpage

\end{document}